\documentclass[conference]{IEEEtran}
\usepackage{color}
\usepackage{extarrows}
\usepackage{amsmath}
\usepackage{amssymb}
\usepackage{proof}
\usepackage{subfig}
\usepackage[all]{xy}
\usepackage{setspace}
\usepackage{url}

\newtheorem{theorem}{\bf Theorem}
\newtheorem{lemma}{\bf Lemma}
\newtheorem{proposition}{\bf Proposition}
\newtheorem{example}{\bf Example}
\newtheorem{definition}{\bf Definition}

\ifCLASSOPTIONcompsoc
  \usepackage[nocompress]{cite}
\else
  \usepackage{cite}
\fi

\ifCLASSINFOpdf
\else
\fi

\begin{document}
\title{Modeling and Reasoning About Wireless Networks: A Graph-based Calculus Approach}

\author{\IEEEauthorblockN{Shichao Liu}
\IEEEauthorblockA{State Key Laboratory of Computer Science,\\ Institute of Software,
 Chinese Academy of Sciences,\\ Beijing, China\\
University of Chinese Academy of Sciences,
Beijing, China\\
Email: liusc@ios.ac.cn}
\and
\IEEEauthorblockN{Ying Jiang}
\IEEEauthorblockA{State Key Laboratory of Computer Science,\\ Institute of Software,
 Chinese Academy of Sciences,\\ Beijing, China\\
Email: jy@ios.ac.cn}
}


% compsoc conferences compared to traditional conferences), use this
%Telephone: (800) 555--1212, Fax: (888) 555--1212}

\maketitle
\begin{abstract}
We propose a graph-based process calculus for modeling and reasoning about wireless networks with local broadcasts.
Graphs are used at syntactical level to describe the topological structures of networks.
This calculus is equipped with a reduction semantics and a labelled transition semantics.
The former is used to define weak barbed congruence.
The latter is used to define a parameterized weak bisimulation emphasizing locations and local broadcasts.
We prove that weak bisimilarity implies weak barbed congruence.
The potential applications are illustrated by some examples and two case studies.
\end{abstract}

\IEEEpeerreviewmaketitle

\section{Introduction}
With the widespread use of wireless communication devices, wireless networks are becoming more important in various fields.
%Classical applications of wireless networks include ambient intelligence, wireless local area networks and wireless sensor networks, etc.
In wireless networks, local broadcast is one of the most important features. Messages, which are transmitted in a limited area, can only be received by the devices linked with the transmitter in wireless networks.
How to ensure that wireless networks can behave in a reasonable manner becomes a critical issue. Assuring the correctness of behaviours of wireless systems is also a difficult task.
The goal of this paper is to develop a graph-based calculus to model and reason about wireless networks.

Formalization of wireless systems has attracted the attentions of many researchers.
Process calculi, e.g. Milner's CCS \cite{Milner1989} and Hoare's CSP \cite{Hoare1985communicating},
provide a good framework to study concurrent systems in a point-to-point approach.
Process calculi for broadcast systems were first studied by Prasad in the work of a calculus for broadcast systems (CBS) \cite{prasad1995calculus}.
However, the broadcast is global in CBS, i.e. messages can be received by all devices in the systems.

%Assuring the correctness of behaviours of wireless systems is a difficult task. Formal methods have a great potential in addressing this problem.
Indeed, local broadcast is a challenge for researchers.
Recently, several process calculi have been proposed to study wireless systems, e.g. \cite{fehnker2012process,godskesen2007calculus,lanese2010operational,merro2009observational,nanz2006framework,singh2010process}.
These process calculi deal with local broadcasts typically by carrying separate topological structures, e.g. using locations and transmission radii for nodes at the syntactic level \cite{lanese2010operational,merro2009observational}, or indexing labelled transitions by graphs at the semantic level \cite{nanz2006framework}.
Behavioural equivalences, e.g. bisimulation, are important tools in \cite{merro2009observational,godskesen2007calculus,nanz2006framework} to reason about wireless systems.
However, these behavioural equivalences do not take location information and links into account,
and, indeed, they straightly use the identity relation of locations to relate actions (i.e. the challenger and the responser must play the same actions at the same locations).
Therefore, these behavioural equivalences cannot relate implementations and specifications with different number of nodes (see Example \ref{example_bisim}).
As pointed out by Lanese and Sangiorgi \cite{lanese2010operational}, {\it ``in wireless systems, each device -- and therefore presumably also the observer -- has a location and a transmission cell, and it is not clear how to take them into account".}
%
%Though behavioural equivalences have been proposed and studied in \cite{merro2009observational,godskesen2007calculus,nanz2006framework},
%to our best knowledge, they do not take location information and links into account,
%and, indeed, they straightly used the identity relation of locations when define behavioural equivalences (i.e. both the challenger and the responser play the same actions at the same locations).
In fact, providing a process calculus with behavioural equivalences by taking locations and links (i.e. transmission cells \cite{lanese2010operational}) into account is non-trivial. This paper makes an effort for this by using graphs, which can concisely specify locations and links of networks.

%In particular the development of an observational theory for establishing process identities is not sufficiently studied in the literature.

In this paper, we propose a Graph-based Calculus for Wireless Networks (called GCWN)
and the semantics of GCWN are suitable for wireless links and local broadcasts.
Behavioural equivalences of GCWN are also well studied by taking locations and links into account.
Graphs play an important role in GCWN, and make it easy to specify and reason about local broadcasts in wireless networks.
%To the best of our knowledge, GCWN is the first behavioral theory for wireless systems by taking locations and links into account.
We hope that the method in this paper can be applied to other process calculi for wireless systems.
This paper makes the following contributions.

{Firstly,} {a graph-based calculus for wireless networks is proposed, where the topology of a wireless network is specified by a graph at the syntactic level.}
In a network, vertices of the associated graph are locations for processes (i.e. nodes) and edges of the graph represent the connections between nodes.

%In a wireless network, a node can proceed by using channels to broadcast messages, and other nodes can receive the messages if there is an edge between the transmitter and the receiver.

{Secondly,} in order to capture evolutions of wireless networks, 
%When developing the semantics, we adhere to the standard operational semantics of typical process calculi, e.g. CCS \cite{Milner1989}, broadcast systems \cite{merro2009observational,godskesen2007calculus,nanz2006framework,lanese2010operational}  and processes with locations \cite{boudol1994theory}.
we define both a reduction semantics of the form $M\xrightarrow[]{}M^{\prime}$, and a labelled transition semantics of the form $M\xrightarrow[]{p:\alpha}M^{\prime}$ (observable transition performing action $\alpha$ at location $p$) or $M\xrightarrow[]{\tau}M^{\prime}$ (unobservable transition).
%Graphs play an important role in the two kinds of semantics.
%In reduction transitions, a node can broadcast a message and all nodes connected with it can receive the message. In labelled transitions, a node can broadcast or receive a message.
As the first theoretical result, we prove that the two semantics describe the same behaviours. % (see Theorem \ref{harmony}).
%Similar results can also be found in \cite{lanese2010operational,merro2009observational,wu2015calculus,lanotte2016semantic}.

{Thirdly,} {two kinds of behavioural equivalences for GCWN are developed.}
%Behavioural equivalence is a central idea in process calculi.
%The main goal of this paper is to develop a sound and complete behavioral theory of wireless networks with explicit locations visible to observers (or environments).
We first adopt the concept of {\it barb} \cite{Milner1992barbed} to define a weak barbed congruence without location information. Barbed congruence is natural to describe that two networks are identical if they exhibit the same barbs during their reductions in any context.
However, barbed congruence is hard to handle directly, because one has to consider all possible contexts by the definition.
Instead, labelled transition systems (LTSs) are widely adopted to study behavioural equivalences. LTSs derive the concept of bisimulation, which is more tractable and equipped with powerful proof techniques.
We define a weak bisimulation for GCWN by taking locations and links into account.
%We define two kinds of notions to characterize when wireless networks (or systems) are indistinguishable.
As another theoretical result, we prove that weak bisimilarity implies weak barbed congruence, i.e. {\it soundness}. %, and vice versa, i.e. {\it completeness}.

{Last,} the potential applications of GCWN are illustrate by examples and case studies. Specially, we use GCWN to model and reason about scenarios in protocol ARAN \cite{sanzgiri2005authenticated} and the Alternating Bit Protocol.

The rest of this paper is organized as follows. Section \ref{sec:syntax} provides the syntax of GCWN. Section \ref{sec:semantics} presents the operational semantics and two kinds of behavioural equivalences for GCWN. Section \ref{sec:main_result} proves that the two semantics coincide and weak bisimilarity implies weak barbed congruence. Section \ref{sec:case_study} provides two case studies. We discuss related work in Section \ref{Related_work}, and make a conclusion in Section \ref{Conclusions}.

For lack of space, most of the proofs are omitted, %but can be found in \cite{liu2017gcwnfull}.
but can be found in the Appendix.

\section{The Calculus}\label{sec:syntax}
In this section, we define the syntax of GCWN.

\noindent{\bf Graphs.} Let ${\sf Loc}$ be a countable set of locations ranged over by $p,q$, etc.
A finite (undirected) graph $G=(|G|,\frown_G)$ consists of a finite set of locations $|G|$ and a set of edges $\frown_G$ which is a binary relation on $|G|$ such that $p\frown_G q$ implies $q\frown_G p$ (symmetric) and $p\neq q$ (no self-loops).
Graphs are used to describe locations and links of networks.
%For $p\in |G|$, the set of adjacent locations of $p$ in $G$ is denoted by ${\sf Adj}(p,G) = \{q\mid p\frown_G q\}$.

Given disjoint sets $E$ and $F$ with $p\in E$, let $E[F/p]=(E-\{p\})\cup F$, by substituting $p$ with $F$.
Let $G$ and $H$ be graphs with $|G|\cap |H|=\emptyset$ and $p\in |G|$.
We define a graph $G[H/p]$ by substituting location $p$ with $H$, and $G[H/p]$ consists of locations $|G[H/p]|=|G|[|H|/p]$ and edges $q\frown_{G[H/p]}r$ with $q\frown_G r$, or $q\frown_H r$, or $q\frown_G p$ and $r\in|H|$, or $r\frown_G p$ and $q\in |H|$.
It will be used when we consider network contexts.

Given two graphs $G$ and $H$ with disjoint locations and $D\subseteq |G|\times|H|$, we define a new graph $K= G \oplus_D H$ such that $|K|=|G|\cup|H|$ and for any $p,q\in|K|$ if $p\frown_G q$ or $p\frown_H q$ or $(p,q)\in D$ then $p\frown_K q$.
%If $D=\emptyset$, we write $G\oplus H $ for $G\oplus_{\emptyset} H$.
The composition of graphs is useful when we define parallel composition of networks.

\noindent{\bf Expressions.} We use $x,y$, etc. for variables, and $v,v_1$, etc. for values that can be transmitted via channels (defined later).
Moreover, values do not include channels. We use $e,e_1$, etc. for arithmetic expressions, which at least include variables and values.
Specially, we use $b,b_1$, etc. for boolean expressions, which at least include $\{\it false,true\}$.
We do not provide a grammar for values and expressions, because they can be constructed with respect to the networks we consider.
The substitutions for expressions are defined as usual denoted by $e\{v/x\}$ and $b\{v/x\}$.
We say that $e$ is data-closed if $e$ does not contain any variable, and similarly for $b$.
We use ${\vec x}$ and ${\vec v}$ for the vectors of variables and values, respectively.

We also assume the existence of an evaluation ${\sf eval}$ for data-closed arithmetic expressions and boolean expressions, returning values and boolean values, respectively.

\noindent{\bf Processes and Networks.}
We define the syntax of GCWN with two levels: a lower one for processes and an upper one for networks.
A network consists of a set of processes, and its topology is specified by a graph.

We use letters $c$, $d$, etc. for channel names, and $\overline{c}$, $\overline{d}$, etc. for co-names.
Let $\mathcal{K}$ be a set of process constants, ranged over by $A$, $B$, etc. For each $A\in \mathcal{K}$, we assume that there is an assigned arity, a non-negative integer, representing the number of parameters that $A$ takes.
The set of all processes, denoted by ${\bf Pr}$, is defined as follows:
$$\small P,Q::={\bf 0}\mid  c(x).P\mid \overline{c}(e).P\mid P + Q \mid {\bf if}~b~{\bf then}~P~{\bf else}~Q \mid A({\vec v})$$
%\begin{center}
%\begin{tabular}{lrrl}
%  % after \\: \hline or \cline{col1-col2} \cline{col3-col4} ...
%  Process~~ &$P,Q$ & $::=$ & ${\bf 0}\mid  c(x).P\mid \overline{c}(e).P\mid{\bf if}~b~{\bf then}~P~{\bf else}~Q \mid A({\vec v})  $\\
%  % after \\: \hline or \cline{col1-col2} \cline{col3-col4} ...
%  Network~~ &$M,N$ & $::=$ & $ G\langle\Phi\rangle \mid M\backslash c \mid M_1\oplus_D M_2  $\\
%\end{tabular}
%\end{center}
%where $f\in \Sigma_n (n \ge 1)$, $x\in {\bf Var}$, $e\in {\bf Exp}$, $b\in {\bf BExp}$, $P_1\ldots,P_n\in {\bf Pr}$, $G$ is a graph, $\Phi$ is a function from $|G|$ to ${\bf Pr}$, $I$ is a finite subset of $\Sigma$, and ${\vec v}$ is consist with the assigned arity of $A$.

Processes are sequential and represent single devices.
${\bf 0}$ is the empty process, meaning a termination.
%Particularly, $\ast$ does not passing any values, and $\ast()\cdot()$ is written as $\ast$ when there is no confusion.
%$\ast$ is an idle process different from the empty sum ${\bf 0}$.
In an input process $c(x).P$, variable $x$ is bound; variables in $e$ are free in an output process $\overline{c}(e).P$.
A sum process $P + Q$ represents a nondeterministic choice.
A conditional process ${\bf if}~b~{\bf then}~P~{\bf else}~Q$ acts as $P$ if $b$ is ${\it true}$, and as $Q$ otherwise, and variables appearing in $b$ are free
in the conditional process.
$A({\vec v})$ denotes a process defined by a (possibly recursive) definition of the form $A({\vec x})\stackrel{\rm def}{=}P$.
The length of ${\vec v}$ and the length of ${\vec x}$ are consistent with the assigned arity of $A$.
A process is {\it data-closed} if all the variables occurring in the process are bound.
The substitution of a value for a variable in processes is denoted by $P\{v/x\}$, which means substituting $v$ for every free occurrence of $x$ in process $P$, and similarly for $P\{{\vec v}/{\vec x}\}$.

%We use $X$, $Y$ for network variables.
The set of all networks, denoted by ${\bf Net}$, is defined as follows:
$$M,N::=G\langle\Phi\rangle \mid M\backslash c \mid M\oplus_D N $$
where, $G$ is a graph and $\Phi$ is a function from $|G|$ to ${\bf Pr}$.

In general, a network is defined by using a graph and a function from locations to processes. The graph specifies the topology of the network and its edges represent the possible communicating capacities between processes. The network $G\langle\Phi\rangle$ is the parallel composition of processes $\Phi(p)\in{\bf Pr}$ for each $p\in |G|$ with communication capabilities specified by $\frown_G$.
Process $\Phi(p)$ is called a node of the network $G\langle\Phi\rangle$. In $G\langle\Phi\rangle$, $\Phi(p)$ and $\Phi(q)$ cannot communicate unless there is an edge between $p$ and $q$.
$M\backslash c$ is a channel restriction ($c$ is bound in $M$), and $c$ is private to $M$.
%The restriction operator $M\backslash c$ has the same meaning as the one in CCS \cite{Milner1989}.
We write $M\backslash I$ as an abbreviation for $M\backslash c_1\ldots\backslash c_k$, with $I=\{c_1,\ldots,c_k\}$.
Moreover, $\alpha$-conversion on channels is defined as usual.

$M\oplus_D N$ represents that two networks can be composed as a new network.
Given $M=G\langle\Phi\rangle\backslash I$ ($I$ can be $\emptyset$) and $N=H\langle\Psi\rangle\backslash J$ ($J$ can be $\emptyset$) with $|G|\cap |H| = \emptyset$, $D\subseteq |G|\times|H|$ and $I\cap J =\emptyset$ (always possible by $\alpha$-conversion on channels), we define the network $M\oplus_D N$ as
$(G\oplus_D H)\langle\Phi^{\prime}\rangle\backslash(I\cup J)$ such that $\Phi^{\prime}(p) = \Phi(p)$ if $p\in |G|$ and $\Phi^{\prime}(p) = \Psi(p)$ if $p\in |H|$.
When $D$ is empty, we write it as $M\oplus N$ for simplicity.
%More generally, $\oplus{\vec P}$ stands for  $P_1\oplus\cdots\oplus P_n$ when ${\vec P} = (P_1,\ldots,P_n)$.
The network $M\oplus_D N$ can be written as $M\mid N$, if $D=|G|\times|H|$.

%{\color{blue}
%Given networks $M=G\langle\Phi\rangle\backslash I$ and $N=H\langle\Psi\rangle\backslash J$ with $p\in |G|$ and $|G|\cap |H|=\emptyset$ and $I\cap J = \emptyset$, $M[N/p]$ represents the network $G[H/p]\langle\Phi^{\prime}\rangle\backslash(I\cup J)$ with $\Phi^{\prime}(p^{\prime}) = \Phi(p^{\prime})$ for $p^{\prime}\notin |H|$ and $\Phi^{\prime}(p^{\prime}) = \Psi(p^{\prime})$ for $p^{\prime}\in |H|$. In general, a substitution may require $\alpha$-conversions on both data variables and channels.
%}

If $M= G\langle\Phi\rangle\backslash c$ (or $M= G\langle\Phi\rangle$), we denote $|M| = |G|$. Meanwhile, for $p\in |G|$, let $M(p)$ represent $\Phi(p)$ and $\frown_M$ represent $\frown_G$. % and we write ${\sf Adj}(p,M)$ instead of ${\sf Adj}(p,G)$.
In this paper, when we talk about several networks together, we implicitly assume that their locations are pairwise disjoint.
A network $M$ is {\it data-closed} if all the variables occurring in $M$ are bound, i.e. every process in $M$ is data-closed.

{\it In the rest of the paper, we focus on data-closed networks.}

\section{Operational Semantics}\label{sec:semantics}
We first define a structural congruence, $\equiv$, as an auxiliary relation to state reduction semantics.
Structural congruence is defined as the congruence satisfying the rules in Table \ref{struct_cong}.
\begin{table}[tp!]\small
  \centering
  \begin{tabular}{ll}
  \hline
  Processes:&\\
  \hline
  $P + {\bf 0}\equiv P$ & \\
  $P + Q \equiv Q + P$ & \\
  $P + (Q + R) \equiv (P + Q) + R$ & \\
  ${\bf if}~ b ~{\bf then}~ P~ {\bf else}~ Q\equiv P$,   &if ${\sf eval}(b)={\it true}$ \\
  ${\bf if}~ b ~{\bf then}~ P~ {\bf else}~ Q\equiv Q$, &if ${\sf eval}(b)={\it false}$ \\
  $A({\vec v})\equiv P\{{\vec v}/{\vec x}\}$,            &if $A(\vec x)\stackrel{\rm def}{=} P$\\
  \hline
  \hline
  Networks:\\
  \hline
  $\alpha$-conversion on channels &\\
  $M\backslash c\backslash d \equiv M\backslash d \backslash c$  & \\
  $ M\oplus_D N\equiv N \oplus_{D} M $&\\
  $ (M\oplus_D N)\backslash c\equiv M \oplus_D (N\backslash c)$, &if $c$ and $\overline{c}$ not in $M$\\
  $G\langle\Phi\rangle \equiv G\langle\Psi\rangle$, &if for all $p\in |G|$, $\Phi(p)\equiv \Psi(p)$ \\
  \hline
\end{tabular}
  \caption{Structural Congruence}\label{struct_cong}
\end{table}

\subsection{Reduction Semantics}
Given a network $M$, a location $p\in|M|$ and a process $Q$, let $M[p\mapsto Q]$ represent a new network obtained from $M$ by
replacing the process at location $p$ by $Q$.
We write $M[p_1\mapsto Q_1][p_2\mapsto Q_2]$ for updating $M$ by replacing processes at $p_1$ and $p_2$ by $Q_1$ and $Q_2$, respectively.
Given a set of locations $L\subseteq |M|$, we write $M[p_i\mapsto Q_i]_{p_i\in L}$ for updating $M$ by changing the process at location $p_i$ into $Q_i$ for each $p_i\in L$.

\begin{figure}[!tp]\small
\setlength{\belowcaptionskip}{-10pt}
  \centering
  $$\infer[\mbox{(R-Bcast)}]{M\xrightarrow[]{}M[p\mapsto P][q_i\mapsto Q_i\{v/x_i\}]_{q_i\in L}}
  {\begin{tabular}{c}
    $M(p)=\overline{c}(e).P + R \qquad {\sf eval}(e) = v $ \\
   $L = \{q_i\mid p\frown_M q_i, M(q_i)=c(x_i).Q_{i}+ R_i\} $ \\
   \end{tabular}
   %M(p)=\overline{c}(e).P + R~~ {\sf eval}(e) = v ~~ L \subseteq {\sf Adj}(p,M)~~\forall q_i\in L. M(q_i)=c(x_i).Q_{i}+ R_i
  }$$
  $$\infer[\mbox{(R-Par)}]{M\oplus_D N\xrightarrow[]{}M^{\prime}\oplus_D N}{M\xrightarrow[]{}M^{\prime}\quad {\sf free}(M,D,N)}$$
  $$\infer[\mbox{(R-Res)}]{M\backslash c \xrightarrow[]{} M^{\prime}\backslash c}{M\xrightarrow[]{} M^{\prime}}\quad
\infer[\mbox{(R-Struct)}]{M \xrightarrow[]{} M^{\prime}}{M\equiv N\quad N\xrightarrow[]{} N^{\prime}\quad N^{\prime}\equiv M^{\prime}}$$
  \caption{Reduction Semantics}\label{reduction_semantics}
\end{figure}

The reduction semantics for networks is defined in Fig. \ref{reduction_semantics} and of the form $M\xrightarrow[]{}M^{\prime}$.
In rule (R-Bcast), the process at location $p$ in the network $M$ broadcasts the value of $e$ via the channel ${c}$ to its adjacent nodes (i.e. all the nodes at $q_i$ with $p\frown_M q_i$).
All the adjacent nodes can receive the message via channel $c$.
While all the other nodes (not in $L$), which are not connected with location $p$, or in which channel $c$ is not available, cannot receive the message.
After the broadcast, the sending node is changed into $P$, the receiving nodes are changed into $Q_i\{v/x\}$, and other nodes are unchanged.
Moreover, when $L$ is empty, the broadcast is lost.
The broadcast does not change the topology of the network.
Rule (R-Par) focuses on the parallel composition of two networks, and it describes the situation that no nodes in one network (i.e. $N$) can receive a broadcast in the other network (i.e. $M$). And the condition ${\sf free}(M,D,N)$ denotes that if nodes in $M$ and $N$ are linked by $D$ then every node in $N$ cannot receive the broadcast messages from $M$.
Rules (R-Res) and (R-Struct) are the standard rules in process calculi, representing restriction reduction and structure reduction, respectively.
Let $\xrightarrow[]{}^{\ast}$ denote the reflexive and transitive closure of $\xrightarrow[]{}$.
%$P\xrightarrow[]{}^\ast P^{\prime}$ if there exists $n \geq 1$ such that $P=P_1$, $P^{\prime}=P_n$, and $P_1\xrightarrow[]{}P_2\cdots P_{n-1}\xrightarrow[]{}P_n$.

\begin{example}\label{example:reduction}
Network $N=(\{1,2,3\},\{(1,2),(1,3)\})\langle\Phi\rangle$, with $\Phi(1)= \overline{c}(0).{\bf 0}$, $\Phi(2) = c(x).{\bf 0}+\overline{d}(1).{\bf 0}$
and $\Phi(3)= c(y).{\bf 0}+ d(x).{\bf 0}$.
%Given $P_1 =\overline{c}(0).{\bf 0}$, $P_2=c(x).{\bf 0}+\overline{d}(1).{\bf 0}$ and $P_3 = c(y).{\bf 0}+ d(x).{\bf 0}$, with $\Phi(i)=P_i$ and $i=1,2,3$.
Therefore,  $\Phi(2)$ and $\Phi(3)$ can hear $\Phi(1)$, but $\Phi(2)$ and $\Phi(3)$ cannot hear each other for lack of links.
Using rule (R-Bcast), we can get $N\xrightarrow[]{}N[1\mapsto {\bf 0}][2\mapsto {\bf 0}][3\mapsto {\bf 0}]$ by a broadcast from node $\Phi(1)$.
\end{example}
\subsection{Weak Barbed Congruence}
What is a proper observation, or {\it barb}, in GCWN? Here, we choose to observe channel communications as in standard process calculi.
To accommodate the ordinary concept of barb \cite{Milner1992barbed}, we abandon location information in the following definition.
Moreover, we only choose broadcasting communications as the barb.
Because, in fact, an observer cannot see whether a node receives a broadcast message, but can detect whether there is a node broadcasting a message by listening.
\begin{definition}[\bf Barb]
Given a network $M\equiv N\backslash I$, we say that $\overline{c}$ is a barb of $M$, written $M\downarrow_{\overline c}$, if there is a location $p\in |M|$ such that $M(p)\equiv \overline{c}(e).P+R$ and $c\notin I$.
%Moreover, if $M\equiv N\backslash I$, we have $c\notin I$ and $\overline{c}\notin I$.
\end{definition}
%\begin{definition}[\bf Barb for Canonical Process]
%Let $f\in \Sigma$ and $P$ be a canonical process. We say that $\overline{f}$ is a barb of $P$ at location $p\in |P|$, written $P\downarrow_{\overline{f}}$, if
%${\sf cs}(P(p))$ is one of forms $\overline{f}(e)\cdot(P_1,\ldots,P_n) + S$, ${\bf if}~b~{\bf then}~\overline{f}(e)\cdot(P_1,\ldots,P_n) + S~{\bf else~}Q$ with ${\sf eval} (b) = true$, or ${\bf if}~b~{\bf then}~Q~{\bf else~}\overline{f}(e)\cdot(P_1,\ldots,P_n) + S$ with ${\sf eval}(b) = false$. Moreover, if $P$ is of the form $Q\backslash I$, we have $f\notin I$.
%\end{definition}
\begin{definition}[\bf Weak Barbed Bisimulation]
A binary relation $\mathcal{B}$ on ${\bf Net}$ is a weak barbed bisimulation if it is symmetric and whenever $(M,N)\in \mathcal{B}$ the following conditions hold:
\begin{itemize}
  \item $M\xrightarrow[]{} M^{\prime}$ implies $N\xrightarrow[]{}^\ast N^{\prime}$ and $(M^{\prime}, N^{\prime})\in\mathcal{B}$ for some $N^{\prime}$;
  \item $M\downarrow_{\overline c}$ implies $N\xrightarrow[]{}^{\ast}N^{\prime}$ and $N^{\prime}\downarrow_{\overline c}$ for some $N^{\prime}$.
\end{itemize}
Weak barbed bisimilarity, denoted by $\stackrel{\bullet}{\approx}$, is the union of all weak barbed bisimulations.
\end{definition}

\begin{example}
Recall the network $N$ in Example \ref{example:reduction}. We have $N\downarrow_{\overline c}$ and $N\downarrow_{\overline d}$. If we build a network $M=N\backslash c$, we only have $M\downarrow_{\overline d}$. Because, the message from channel ${c}$ has been restricted and cannot be observed by environments.
\end{example}

\begin{lemma}\label{lemma:barbbisim}
$\stackrel{\bullet}{\approx}$ is an equivalence relation.
\end{lemma}
%\begin{IEEEproof}
%We need to prove that $\stackrel{\bullet}{\approx}$ is reflexive, symmetric and transitive. It is straightforward from the definition.
%\end{IEEEproof}

%We define a special set of network contexts with the assumption that all the nodes in an observer (i.e. a context) are linked with the nodes in the network filled in the context because weak barbed bisimulation does not take locations into account.
%
%\begin{definition}
%Network contexts are networks with one hole $[\cdot]$, defined by
%$$\mathcal{C}[\cdot]::= [\cdot]~~\mid~~ M\mid [\cdot]~~\mid~~ [\cdot]\mid M ~~\mid~~ [\cdot]\backslash c.$$
%\end{definition}

\begin{definition}
%A context $\mathcal{C}[\cdot]$ is a network term with one hole, denoted by $[\cdot]$.
Network contexts are networks with one hole $[\cdot]$ (i.e. a special node with a location $p$), defined by
$$\mathcal{C}[\cdot]::= [\cdot]\mid M\oplus_D [\cdot]\mid [\cdot]\oplus_D M \mid [\cdot]\backslash c$$
%$$\mathcal{C}[\cdot]::= [\cdot]\mid (M\mid [\cdot])\mid ([\cdot]\mid M) \mid [\cdot]\backslash c.$$
where edges $D\subseteq \{(p,q)\mid q\in|M|, p \mbox{ is the location of }[\cdot]\}$.
\end{definition}

$\mathcal{C}[N]$ means putting the network $N$ into the hole at location $p$.
Every edge $(p,q)\in D$ will be replaced by edges $(p,r)$ with $r\in |N|$.
For instance, let $\mathcal{C}[\cdot] = G\langle\Phi\rangle\backslash I$, $N=H\langle\Psi\rangle\backslash J$, $I\cap J=\emptyset$ and assume that the hole's location is $p$, then $\mathcal{C}[N]$ represents the network $G[H/p]\langle\Phi^{\prime}\rangle\backslash(I\cup J)$ such that $\Phi^{\prime}(q)=\Phi(q)$ if $q\notin |H|$ and $\Phi^{\prime}(q) = \Psi(q)$ if $q\in|H|$. Here $I$ can be empty, and similarly for $J$.

%$\mathcal{C}[N]$ means putting the network $N$ into the hole $[\cdot]$.
%%For instance, if $\mathcal{C}[\cdot] = M\mid [\cdot]$, then $\mathcal{C}[N]= M\mid N$.
%%Because the barbed bisimulation does not take locations into account, we define a special set of network contexts with the assumption that all the nodes in an observer (i.e. a context) are linked with the nodes in a network filled in the context.
%For instance, context $ M\mid [\cdot]$ means that we can put a network $N$ into the hole and obtain a new network $M\mid N$ by adding the connections between all nodes in $M$ and all nodes in $N$.

\begin{proposition}\label{prop:congruence}
For any equivalence relation $\mathcal{R}\subseteq {\bf Net}\times{\bf Net}$, there exists a largest congruence $\overline{\mathcal{R}}$ contained in $\mathcal{R}$. This relation is characterized by $(M,N)\in \overline{\mathcal{R}}$ if and only if for any context $\mathcal{C}[\cdot]$ one has
$(\mathcal{C}[M],\mathcal{C}[N])\in \mathcal{R}$.
\end{proposition}
%\begin{IEEEproof}For the first statement,
%from the definition of congruence, it is obvious that the identity relation contained in $\mathcal{R}$ is a congruence. And congruences are closed under arbitrary unions and contexts.
%
%For the second statement, let $\mathcal{E}$ be a congruence defined by  $(M,N)\in \mathcal{E}$ if and only if for any context $\mathcal{C}[\cdot]$ one has $(\mathcal{C}[M],\mathcal{C}[N])\in \mathcal{R}$. Therefore, $\mathcal{E}$ is a congruence contained in $\mathcal{R}$ (because we can take $\mathcal{C}[\cdot] = [\cdot]$) and hence $\mathcal{E}\subseteq \overline{\mathcal{R}}$. Conversely, let $(M,N)\in \overline{\mathcal{R}}$ and $\mathcal{C}[\cdot]$ be a context. Because $\overline{\mathcal{R}}$ is a congruence, we have $(\mathcal{C}[M],\mathcal{C}[N])\in \overline{\mathcal{R}}$. We have $(\mathcal{C}[M],\mathcal{C}[N])\in \mathcal{R}$ from $\overline{\mathcal{R}}\subseteq \mathcal{R}$ by definition of $\overline{\mathcal{R}}$ and hence $(M,N)\in \mathcal{E}$.
%\end{IEEEproof}

\begin{definition}
Networks $M$ and $N$ are weakly barbed congruent, denoted by $M \cong N$, if $\mathcal{C}[M]\stackrel{\bullet}{\approx}\mathcal{C}[N]$ for any context $\mathcal{C}[\cdot]$.
\end{definition}

$\cong$ is the largest congruence in $\stackrel{\bullet}{\approx}$ by Proposition \ref{prop:congruence}.

\subsection{Labelled Transition Semantics}
In this paper, the labelled transition systems of GCWN are divided into two parts: one part for processes and the other part for networks.

Fig. \ref{LTS_process} describes the labelled transition semantics for processes.
The transitions are of the form $P\xrightarrow[]{\alpha}P^{\prime}$, and the syntax of action $\alpha$ is defined as
$$\alpha::= cv\mid \overline{c}v$$
where action $cv$ stands for receiving a broadcast message $v$ via channel $c$ and action $\overline{c}v$ stands for broadcasting message $v$ via channel ${c}$. The rules in Fig. \ref{LTS_process} are self-explanatory.

\begin{figure}[!tp]\small
  \centering
  $$\infer[\mbox{(P-Output)}]{\overline{c}(e).P\xrightarrow[]{\overline{c}v}P}{{\sf eval}(e)=v}\quad
  \infer[\mbox{(P-Input)}]{c(x).P\xrightarrow[]{cv}P\{v/x\}}{}$$
  $$\infer[\mbox{(P-SumL)}]{P+Q\xrightarrow[]{\alpha}P^{\prime}}{P\xrightarrow[]{\alpha}P^{\prime}}\quad
  \infer[\mbox{(P-IfT)}]{{\bf if}~ b ~{\bf then}~ P~ {\bf else}~ Q\xrightarrow[]{\alpha}P^{\prime}}{{\sf eval}(b)={\it true}\quad P\xrightarrow[]{\alpha}P^{\prime}}
  $$
  $$\infer[\mbox{(P-SumR)}]{P+Q\xrightarrow[]{\alpha}Q^{\prime}}{Q\xrightarrow[]{\alpha}Q^{\prime}}\quad
  \infer[\mbox{(P-IfF)}]{{\bf if}~ b ~{\bf then}~ P~ {\bf else}~ Q\xrightarrow[]{\alpha}Q^{\prime}}{{\sf eval}(b)={\it false}\quad Q\xrightarrow[]{\alpha}Q^{\prime}}$$
  $$\infer[\mbox{(P-Rec)}]{A({\vec v})\xrightarrow[]{\alpha}P^{\prime}}{P\{{\vec v}/{\vec x}\}\xrightarrow[]{\alpha}P^{\prime}\quad A({\vec x})\stackrel{\rm def}{=}P}$$
  \caption{Labelled Transition Semantics for Processes}\label{LTS_process}
\end{figure}

Fig. \ref{LTS_network} describes the labelled transition semantics for networks.
The transitions for networks are of the form $M\xrightarrow[]{\delta}M^{\prime}$. The grammar for $\delta$ is
$$\delta::= p:\alpha \mid \tau$$
where $p\in {\sf Loc}$ and $\alpha$ is an action.
$\xrightarrow[]{p:\alpha}$ represents that the node at location $p$ performs an action $\alpha$.
%broadcasts a message $v$ via co-channel $\overline{c}$.
%Similarly, $\xrightarrow[]{p:cv}$ represents that a node located at $p$ receives a broadcasting message via channel $c$.
And $\xrightarrow[]{\tau}$ represents the unobservable transition.
\begin{figure}[!tp]\small
  \centering
  $$\infer[\mbox{(N-Send)}]{M\xrightarrow[]{p:\overline{c}v}M[p\mapsto P^{\prime}]}{p\in |M|\quad M(p)=P\quad P\xrightarrow[]{\overline{c}v} P^{\prime}}$$
  $$\infer[\mbox{(N-Recv)}]{M\xrightarrow[]{p:cv}M[p\mapsto P^{\prime}]}{p\in |M|\quad M(p)=P\quad P\xrightarrow[]{cv} P^{\prime}}$$
$$\infer[\mbox{(N-Bcast)}]
{ \begin{tabular}{c}
      $M\oplus_D N\xrightarrow[]{p:\overline{c}v}M^{\prime}\oplus_{D}N^{\prime}$\\
      $N\oplus_D M\xrightarrow[]{p:\overline{c}v}N^{\prime}\oplus_{D}M^{\prime}$\\
  \end{tabular}
}
{M\xrightarrow[]{p:\overline{c}v}M^{\prime}\quad N\xrightarrow[]{q:cv}N^{\prime}\quad (p,q)\in D}$$
$$\infer[\mbox{(N-Res1)}]{M\backslash c \xrightarrow[]{\tau}M^{\prime}\backslash c}{M \xrightarrow[]{p:\overline{c}v}M^{\prime}} \quad
\infer[\mbox{(N-Res2)}]{M\backslash c \xrightarrow[]{\delta}M^{\prime}\backslash c}{M \xrightarrow[]{\delta}M^{\prime}\quad c\mbox{ and }\overline{c}\mbox{ not in } \delta}$$
$$\infer[\mbox{(N-ParL)}]{M\oplus_D N\xrightarrow[]{\delta}M^{\prime}\oplus_{D}N}{M\xrightarrow[]{\delta}M^{\prime}}\quad
\infer[\mbox{(N-ParR)}]{M\oplus_D N\xrightarrow[]{\delta}M\oplus_{D}N^{\prime}}{N\xrightarrow[]{\delta}N^{\prime}}$$
  \caption{Labelled Transition Semantics for Networks}\label{LTS_network}
\end{figure}

Rule (N-Send) models the broadcast at location $p$ of value $v$ via channel ${c}$.
Rule (N-Recv) shows that a value can be received at location $p$ via channel $c$.
Rule (N-Bcast) describes the propagation of a broadcast, and the premise $(p,q)\in D$ makes sure that only the nodes connected with the transmitter can receive the message.
Rule (N-Res1) hides a broadcast in restricted networks.
Rule (N-Res2) is standard in process calculi.
Rules (N-ParL) and (N-ParR) model parallel composition networks.
%{\color{blue}Rule (N-ParL) models parallel composition networks, moreover, when $\delta$ is of the form $p:\overline{c}v$ rule (N-Par) models the situation where potential receivers cannot receive the message for lack of connections, and similarly for rule (N-ParR).
%}

We write $M \xrightarrow[]{\tau^{\ast}}M^{\prime}$ if there exists
$n\geq 1$ such that $M = M_1$, $M^{\prime} = M_n$, $M_1\xrightarrow[]{\tau}M_2\xrightarrow[]{\tau}\cdots \xrightarrow[]{\tau}M_n$.
$M\xLongrightarrow[]{p:\alpha}M^{\prime}$ denotes $M\xrightarrow[]{\tau^{\ast}}M_1\xrightarrow[] {p:\alpha}M_1^{\prime}\xrightarrow[]{\tau^{\ast}}M^{\prime}$ for some $M_1$ and $M_1^{\prime}$.

%Now we give some examples to illustrate two broadcast mechanisms, i.e. one is the usual broadcast and the other is the point-to-point communication using tree structures.
\begin{example}[Broadcast]
Let $N_i=(\{i\},\emptyset)\langle\Phi_i\rangle$, $i=1,2,3$, be networks, i.e. each network only has one node.
Let $\Phi_1(1) =\overline{c}(0).{\bf 0}$, $\Phi_2(2)=c(x).{\bf 0}$ and $\Phi_3(3) = c(y).{\bf 0}$.
%and the networks $N_i=(\{i\},\emptyset)\langle\Phi_i\rangle$ with $\Phi_i(i)=P_i$.
Using the rules in Fig. \ref{LTS_process} and Fig. \ref{LTS_network},
$N_1\mid (N_2\oplus N_3)$ can evolve as follows:
$$\tiny \infer{N_1\mid (N_2\oplus N_3)\xrightarrow[]{1:\overline{c}0}N[1\mapsto{\bf 0}]\mid (N_2[2\mapsto{\bf 0}]\oplus N_3[3\mapsto{\bf 0}])}
{\infer{N_3\xrightarrow[]{3:c0}N_3[3\mapsto {\bf 0}]}{N_3(3)=c(y).{\bf 0}\quad c(y).{\bf 0}\xrightarrow[]{c0}{\bf 0}} \qquad
 \infer{N_1\mid N_2\xrightarrow[]{1:\overline{c}0}N_1[1\mapsto{\bf 0}]\mid N_2[2\mapsto{\bf 0}]}
       {\infer{N_1\xrightarrow[]{1:\overline{c}0}N_1[1\mapsto {\bf 0}]}
              {\begin{tabular}{c}
                 $N_1(1)=\overline{c}(0).{\bf 0}$ \\
                 $\overline{c}(0).{\bf 0}\xrightarrow[]{\overline{c}0}{\bf 0}$ \\
               \end{tabular}
               }\qquad
        \infer{N_2\xrightarrow[]{2:c0}N_2[2\mapsto {\bf 0}]}
              {\begin{tabular}{c}
                 $ N_2(2)=c(x).{\bf 0}$ \\
                 $c(x).{\bf 0}\xrightarrow[]{c0}{\bf 0}$ \\
               \end{tabular}
              }}}$$
\end{example}

\subsection{Weak Bisimulation}
%In fact, the connectivity of wireless networks can no longer be abstracted into a simple topology, e.g. a complete graph.
As explained in the introduction, observers (or environments) should take the links and locations into account.
We define a weak bisimulation for GCWN to take them into account,
%Therefore, this behavioral equivalence is defined with respect to that an observer can interact with the networks.
and it makes observers more context-sensitive when the observers interact with networks.

For instance, when networks $M=G\langle\Phi\rangle$ and $N=H\langle\Psi\rangle$ are bisimilar, the observer has to point out which node $\Phi(p)$ of $M$ and which node $\Psi(q)$ of $N$ should be related in the bisimulation relation.
%Therefore, when considering whether two systems (or processes) are equal, one should consider the devices (subprocesses in a process $P$) in one network (i.e. process $P$) and the devices in the other network pairwise.
Inspired by \cite{Ehrhard2013ccts,liu2016vccts},
we define a localized relation on ${\bf Net}$ through triples $(M,E,N)$ by taking locations into account, and $E \subseteq |M|\times|N|$ specifies the pairs of locations of $M$ and $N$. Let $E^{-1}=\{(q,p)\mid (p,q)\in E\}$.

\begin{definition}[\bf Localized Relation]
A localized relation on ${\bf Net}$ is a set $\mathcal{R}\subseteq {\bf Net}\times\mathcal{P}({\sf Loc}^2)\times{\bf Net}$ such that, if $(M,E,N)\in \mathcal{R}$ then $E\subseteq |M|\times|N|$. $\mathcal{R}$ is symmetric if $(P,E,Q)\in \mathcal{R}$ implies $(Q,E^{-1},P)\in \mathcal{R}$.
\end{definition}

\begin{definition}[\bf Weak Bisimulation]
A symmetric localized relation $\mathcal{R}$ on ${\bf Net}$ is a weak bisimulation such that whenever $(M,E,N)\in \mathcal{R}$:
\begin{itemize}
  \item if $M\xlongrightarrow[]{\tau}M^{\prime}$, then there is $N^{\prime}$ such that
  $N\xrightarrow[]{\tau^{\ast}}N^{\prime}$ and $(M^{\prime},E,N^{\prime})\in\mathcal{R}$;
  \item if $M\xlongrightarrow[]{p:\alpha}M^{\prime}$, then  there is $N^{\prime}$ such that $N\xLongrightarrow[]
  {q:\alpha}N^{\prime}$ with $(p,q)\in E$, and $(M^{\prime},E,N^{\prime})\in \mathcal{R}$.
\end{itemize}
%The weak bisimilarity is the union of all weak bisimulations, denoted by $\approx$.
\end{definition}
\begin{definition}
$M$ and $N$ are weakly bisimilar, denoted by $M\approx N$, if there exist a weak bisimulation $\mathcal{R}$ and a relation $E\subseteq |M|\times |N|$ such that $(M,E,N)\in \mathcal{R}$.
\end{definition}

In the definitions, $E$ can be taken as a parameter and the weak bisimulation can be called parameterized weak bisimulation, similar to the parameterized location bisimulation in \cite{boudol1994theory}. Moreover, if $E=|M|\times|N|$, we obtain a weak bisimulation ignoring the location information.
%\begin{lemma}
%$\approx$ is a weak bisimulation.
%\end{lemma}

%\begin{example}[Comparisons on Bisimulations]
%Given two networks $M$ and $N$ such that each network only has one node, i.e. $M=(\{p\},\emptyset)\langle\Phi\rangle$ and $N=(\{q\},\emptyset)\langle\Psi\rangle$ for some locations $p$ and $q$, and some functions from locations to processes $\Phi$ and $\Psi$. Then for any process $P$, we have $M[p\mapsto P]\approx N[q\mapsto P]$, by showing that the localized relation
%$$\{(M[p\mapsto P],\{(p,q)\},N[q\mapsto P])\mid \mbox{ for any }P\}$$
%is a weak bisimulation. This is different from other weak bisimulations in the literature \cite{merro2009observational,godskesen2007calculus,kouzapas2011process},
%where weakly bisimilar networks must play the same observable actions on the same locations.
%For instance, Theorem 6.1 in \cite{merro2009observational} makes use of rule (Move), which allows a mobile node changing its location, to make two mobile nodes with
%the same process weakly bisimilar. But the method does not work for stationary nodes (i.e. nodes that never change their locations).
%\end{example}

\begin{example}[Comparisons on Bisimulations]\label{example_bisim}
In GCWN, we define a network $Sys$ for a simple protocol, transferring data from one node to another. We provide a network ${ Spec}$ as a specification for the protocol.
We define the sender $P$ and the receiver $Q$ in $Sys$, and
the process $R$ in $Spec$ as follows
\begin{center}
\begin{tabular}{rcl}
  $P$ & $\stackrel{\rm def}{=}$ & $\overline{c_1}(0).\overline{d_1}(1).d_2(x).P$ \\
  $Q$ & $\stackrel{\rm def}{=}$ & $d_1(x).\overline{c_2}(0).\overline{d_2}(1).Q$ \\
  %${ Sys}$ & $\stackrel{\rm def}{=}$  & $G\langle\Phi\rangle\backslash \{d_1,d_2\}$ \\
  $R$ & $\stackrel{\rm def}{=}$ & $\overline{c_1}(0).\overline{c_2}(0).R$ \\
\end{tabular}
\end{center}
Channels $d_1$ and $d_2$ are used to transfer data and acknowledgements between $P$ and $Q$.
Let $Sys=G_1\langle\Phi\rangle\backslash \{d_1,d_2\}$ and $Spec=G_2\langle\Psi\rangle$,
where $G_1=(\{1,2\},\{(1,2)\})$, $\Phi(1) = P$, $\Phi(2)= Q$, $G_2=(\{3\},\emptyset)$ and $\Psi(3)= R$.
%For $Spec=G_2\langle\Psi\rangle$, its associated graph is $G_2=(\{3\},\emptyset)$.
From $({ Sys},\{(1,3),(2,3)\},{ Spec})$, we can build a weak bisimulation containing it in GCWN, i.e. ${ Sys}\approx { Spec}$.
However, $Sys$ and $Spec$ are not weakly bisimilar in the literature \cite{merro2009observational,godskesen2007calculus,nanz2006framework},
where weakly bisimilar networks must play the same observable actions at the same locations.
\end{example}

\section{Main Result}\label{sec:main_result}
In this section, we show that reduction semantics and labelled transition semantics model the same behaviours,
and prove that weak bisimilarity implies weak barbed congruence.
\subsection{Harmony Theorem}
We have defined reduction semantics and labelled transition semantics for networks in the previous section.
There is a close relation between them, i.e. the internal reduction and the labelled transition describe the same behaviours. %, up to structural congruence $\equiv$.
Before proving this, we provide two lemmas.
\begin{lemma}\label{equiv_action}
{~
\begin{enumerate}
  \item If $M\xrightarrow[]{p:cv}M^{\prime}$, then there are $v$, $x$, $I$ with $c\notin I$ and $p\in |M|$ such that $M\equiv  N\backslash I$, $N(p)=c(x).P+R$ and $M^{\prime}\equiv N[p\mapsto P\{v/x\}]\backslash I$.
  \item If $M\xrightarrow[]{p:\overline{c}v}M^{\prime}$, then there are $e$ with ${\sf eval}(e) = v$, $I$ with $c\notin I$, $M(p)=\overline{c}(e).P + R$, $L=\{q_i\mid p\frown_M q_i, M(q_i)= c(x_i).Q_i+R_i\}$, $|M_1|=L\cup\{p\}$, $|M_2|=|M|\setminus |M_1|$ and $D\subseteq |M_1|\times|M_2|$ consistent with $\frown_M$ such that
      $M\equiv (M_1\oplus_D M_2)\backslash I$, and $M^{\prime}\equiv (M_1[p\mapsto P][q_i\mapsto Q_i\{v/x\}]_{q_i\in L} \oplus_D M_2)\backslash I$.
\end{enumerate}
}
\end{lemma}
\begin{IEEEproof}
Induction on the transition rules for networks.
\end{IEEEproof}

In the second part of Lemma \ref{equiv_action}, we can divide the network $M$ into two parts. One part $M_1$ consists of the broadcasting node at location $p$ and all its adjacent nodes that can receive the message via channel $c$. The other part $M_2$ consists of the nodes that are not connected with location $p$, and the nodes that are connected with $p$ but in which $c$ is not available. %, will not receive the broadcast message.
%All its adjacent nodes can receive the message via channel $c$. While all the other nodes that are not connected with location $p$, or $c$ is not the available, will not receive the broadcast message.
%We need to show that $\equiv_c$ respect the transitions of Fig. \ref{LTS1}.
\begin{lemma}\label{lamma:equiv}
If $M\xrightarrow[]{\delta}M^{\prime}$ and $M\equiv N$, then there exists $N^{\prime}$ such that $N\xrightarrow[]{\delta}N^{\prime}$ and $M^{\prime}\equiv N^{\prime}$.
\end{lemma}

By Lemmas \ref{equiv_action} and \ref{lamma:equiv}, we can prove the following theorem.
\begin{theorem}[Harmony Theorem]\label{harmony}
{
\begin{itemize}
  \item If $M\xrightarrow[]{}M^{\prime}$, then
        \begin{itemize}
          \item either $M\xrightarrow[]{\tau}M^{\prime\prime}$ and $M^{\prime\prime}\equiv M^{\prime}$ for some $M^{\prime\prime}$;
          \item or there is $p:\overline{c}v$ such that $M\xrightarrow[]{p:\overline{c}v}M^{\prime\prime}$ and $M^{\prime\prime}\equiv M^{\prime}$ for some $M^{\prime\prime}$.
        \end{itemize}
  \item If $M\xrightarrow[]{p:\overline{c}v}M^{\prime}$ or $M\xrightarrow[]{\tau}M^{\prime}$, then $M\xrightarrow[]{}M^{\prime}$.
\end{itemize}
}
\end{theorem}
\subsection{Soundness}
%\begin{proposition}
%$\approx$ is an equivalence relation.
%\end{proposition}

We first prove that weak bisimilarity is an equivalence relation.
\begin{lemma}\label{lemma:bisim_equiv}
$\approx$ is an equivalence relation.
\end{lemma}
%\begin{IEEEproof}
%%$\approx$ is symmetric by definition. We need to prove $\approx$ is reflexive and transitive, see Appendix for details.
%Because $\approx$ is reflexive by Lemma \ref{reflexivity} , symmetric from the definition and transitive by Lemma \ref{transitivity}.
%\end{IEEEproof}

Then we prove that $\approx$ is preserved by the operators in networks.
In CCS \cite{Milner1989}, if $\mathcal{R}$ is a weak bisimulation and $P~\mathcal{R}~Q$, then one can prove that, for any $S$, $S\mid P$ and $S\mid Q$ are weak bisimilar, by showing that a new relation $\mathcal{R}^{\prime}$ extending $\mathcal{R}$ (i.e. $(S\mid P)~\mathcal{R}^{\prime}~(S\mid Q)$) is a weak bisimulation.
However, we cannot simply do this in GCWN, because we need to record locations of nodes and links between nodes.
The main challenge is to extend a localized relation $\mathcal{R}$ to another localized relation $\mathcal{R}^{\prime}$
to accommodate the parallel composition in GCWN.
For instance, let $O\oplus_C M$ be a parallel composition of networks $O$ and $M$ with some $C\subseteq|O|\times|M|$. We have to build $O\oplus_D N$ as a parallel composition of networks $O$ and $N$ with some relation $D\subseteq |O|\times|N|$. Meanwhile, the relations $C$ and $D$ should satisfy some constraints.
\begin{definition}[\bf Adapted Triple of Relations]
We say that a triple of relations $(D,D^{\prime},E)$ with $D\subseteq A\times B$, $D^{\prime} \subseteq A\times B^{\prime}$ and $E\subseteq B\times B^{\prime}$ is {\it adapted}, if for any $(a,b,b^{\prime})\in A\times B\times B^{\prime}$ with $(b,b^{\prime})\in E$, $(a,b)\in D$  if and only if $(a,b^{\prime})\in D^{\prime}$.
\end{definition}
\begin{definition}[\bf Parallel Extension]
Let $\mathcal{R}$ be a localized relation. A localized relation $\mathcal{R}^{\prime}$ is a parallel extension of $\mathcal{R}$, if for any $(U,F,V)\in \mathcal{R}^{\prime}$ the following conditions are satisfied:
\begin{itemize}
  \item there exist a network $O$, a triple $(M,E,N)\in\mathcal{R}$, $C\subseteq |O|\times |M|$, $D\subseteq |O|\times |N|$ such that $U=O\oplus_C M$ and $V=O\oplus_D N$,
  \item $(C,D,E)$ is adapted,
  \item $F$ is the relation $({\rm Id}_{|O|}\cup E)\subseteq |U|\times |V|$, and ${\rm Id}_{|O|}=\{(p,p)\mid p\in |O|\}$.
\end{itemize}
\end{definition}

%Here $\rm{Id}_{|O|}=\{(p,p)\mid p\in |O|\}$.
Intuitively, the premise that $(C,D,E)$ is adapted specifies that network $O$ as an observer should have the same connections with $M$ and $N$ up to $E$, i.e. taking links into account.

\begin{proposition}\label{propositionparallel}
If $\mathcal{R}$ is a weak bisimulation, then its parallel extension $\mathcal{R}^{\prime}$ is also a weak bisimulation.
\end{proposition}

Now we can prove the following theorem.
\begin{theorem}\label{bisimilationCongruence}
$\approx$ is a congruence.
\end{theorem}
%\begin{IEEEproof}
%$\approx$ is an equivalence by Lemma \ref{lemma:bisim_equiv}.
%Here, we need to prove that if $M$ and $N$ are two networks, and $M\approx N$, (i.e. $(M,E,N)\in \mathcal{R}$ for some weak bisimulation $\mathcal{R}$), then
%\begin{itemize}
%  \item[(1)] $\mathcal{R}$'s parallel extension is a weak bisimulation,
%  \item[(2)] $(M\backslash c, E, M\backslash c)$ is contained in some weak bisimulation for any channel $c$.
%\end{itemize}
%
%For the proof of (1), we directly apply the Proposition \ref{propositionparallel}.
%
%For the proof of (2), it is sufficient to show that the localized relation
%$$\mathcal{S}\stackrel{\rm def}{=}\{((M\backslash c, E, M\backslash c))\mid (M,E,N)\in \mathcal{R} \mbox{ for any channel } c\}$$
%is a weak bisimulation. Then we do a case analysis on the possible transition $M\backslash c\xrightarrow[]{\delta}M^{\prime}$. The proof is straightforward.
%\end{IEEEproof}

Structurally congruent networks are weakly bisimilar.
\begin{proposition}\label{equiv_bisim}
$M\equiv N$ implies $M\approx N$.
\end{proposition}
\begin{IEEEproof}
Induction on the rules of $\equiv$.
\end{IEEEproof}

Weak bisimulation is reduction closed and barb preserving.
\begin{proposition}\label{propsitionbisimulationbarb}
If $M\approx N$ then $M\stackrel{\bullet}{\approx}N$.
\end{proposition}
%\begin{IEEEproof}
%Let $\mathcal{B}$ be a binary relation on processes defined by: $(M,N)\in \mathcal{B}$ if $M\approx N$. Then we have to prove that $\mathcal{B}$ is a weak bared bisimulation. First, we know that $\mathcal{B}$ is symmetric, because $\approx$ is symmetric. Then we need to prove $\mathcal{B}$ is reduction closed and barb preserving.
%
%(1) Let $(M,N)\in \mathcal{B}$. If $M\xrightarrow[]{}M^{\prime}$ which is $M\xrightarrow[]{p:\overline{c}v}M^{\prime\prime}\equiv M^{\prime}$ by Theorem \ref{harmony}.
%Because $M\approx N$,
%we have $N\xLongrightarrow[]{q:\overline{c}v}N^{\prime}$ (i.e. $N\xrightarrow[]{}^{\ast}N^{\prime}$ by by Theorem \ref{harmony}) and $M^{\prime\prime}\approx N^{\prime}$.
%We also have $M^{\prime\prime}\approx M^{\prime}$ by Proposition \ref{equiv_bisim}. Since $\approx$ is an equivalence, we have $M^{\prime}\approx N^{\prime}$.
%Thus, we have $(M^{\prime},N^{\prime})\in \mathcal{B}$.
%
%(2) Let $(M,N)\in \mathcal{B}$. If $M\downarrow_{\overline c}$,
%then there exists a transition $M\xrightarrow[]{p:\overline{c}v}M^{\prime}$. Since $M\approx N$,
%we have $N\xLongrightarrow[]{q:\overline{c}v}N^{\prime}$ and $M^{\prime}\approx N^{\prime}$.
%$N\xLongrightarrow[]{q:\overline{c}v}N^{\prime}$ means $N\xrightarrow[]{\tau^{\ast}}N_1$ (i.e. $N\xrightarrow[]{}^{\ast}N_1$ by Theorem \ref{harmony})
%with $N_1\downarrow_{\overline c}$ for some $N_1$. From $M\downarrow_{\overline c}$, we get that
%$N\rightarrow^{\ast}N_1$ with $N_1\downarrow_{\overline c}$ as required.
%\end{IEEEproof}

From Proposition \ref{propsitionbisimulationbarb} and Theorem \ref{bisimilationCongruence}, we can easily get the following theorem.
\begin{theorem}[\bf Soundness]\label{soundness}
If $M\approx N$, then $M\cong N$.
\end{theorem}

Though the converse direction of Theorem \ref{soundness} (i.e. completeness) holds in CCS in a point-to-point approach, e.g. \cite{Milner1992barbed,Sangiorgi:2011}, it does not hold in this paper with local broadcast.

\section{Case Studies}\label{sec:case_study}
In this section, we show that GCWN can be used to model and reason about non-trivial networks.
\subsection{ARAN}
We use behavioural equivalences of GCWN to show an attack scenario in ARAN \cite{sanzgiri2005authenticated}.
ARAN is a secure on-demand routing protocol for ad hoc networks. The goal of ARAN is to ensure message integrity and non-repudiation in the route process using public key cryptography.
ARAN requires the use of a trusted server $T$ to send a certification to each node.
Each valid node $A$ in the network has a pair of public and private keys $(K_{A+}, K_{A-})$ and a certification (received from $T$) to authenticate itself to other nodes.
For a node $X$, let $X$'s certification be $cert_{X} = \{ip_{X},K_{X+},t,e\}_{K_{T-}}$, containing IP address of $X$, the public key of $X$, a timestamp $t$ when $cert_X$ was created and a time $e$ at which $cert_X$ expires.
As usual, we use $\{d\}_{K_{X-}}$ to encrypt data $d$ with the private key $K_{X-}$ of node $X$.
To abstract some implementation details,
%To illustrate the protocol precisely,
we introduce some auxiliary functions to manipulate the messages. For instance, we use ${\sf check1}$ and ${\sf check2}$ to check the signature using certifications in the messages at the request and the reply steps, use ${\sf getIP}$ to extract the IP address of the node that broadcasts the message, and use functions ${\sf NewMsg1}$, ${\sf NewMsg2}$ and ${\sf NewMsg3}$ to construct new messages at different steps of the protocol.
We use ${\sf fst}$ (and ${\sf snd}$) to return the first element (and the second element) of a pair.
%We omit the definitions of some functions, but they can be defined easily.

\begin{figure}[!tp]\small
\setlength{\belowcaptionskip}{-10pt}
  \centering
  \begin{tabular}{rl}
  $A(ip_A,ip_X)$  $\stackrel{\rm def}{=}$ & $\overline{c}((\{RDP,ip_X,N_A\}_{K_{A-}},[cert_A])).A_1(ip_A)$  \\
  $A_1(ip_A)$  $\stackrel{\rm def}{=}$ & $d(x).({\bf if}~{\sf check2}(x) = ok ~\wedge~ {\sf snd}(x) = ip_A$ \\
    &\qquad \quad ${\bf then}~\overline{\sf s}(0).{\bf 0}~{\bf else}~A_1(ip_A)) $\\
  $Q(ip_Q)$  $\stackrel{\rm def}{=}$ & $ c(x).({\bf if}~ {\sf check1}(x)= ok ~{\bf then}~$ \\
      &\qquad \quad $  \overline{c}({\sf NewMsg1}(x,ip_Q)).Q_1(ip_Q,{\sf getIP}(x)) $ \\
      &\qquad \quad $ {\bf else}~Q(ip_Q))$\\
  $Q_1(ip_Q,ip)$  $\stackrel{\rm def}{=}$ & $ d(y).({\bf if}~{\sf check2}(y)=ok~ \wedge~ {\sf snd}(y) = ip_Q~{\bf then}~$\\
    & \qquad \quad $\overline{d}(({\sf NewMsg2}(y,ip_Q),ip)).{\bf 0}$\\
    & \qquad \quad ${\bf else}~Q_1(ip_Q,ip))$\\
  $X(ip_X)$ $\stackrel{\rm def}{=}$ & $c(x).({\bf if}~{\sf check1}(x)= ok~{\bf then}~$ \\
   & \qquad \quad $\overline{d}(({\sf NewMsg3}(x,ip_X),{\sf getIP}(x))).{\bf 0} $ \\
   & \qquad \quad ${\bf else}~X(ip_X))$\\
  $I$ $\stackrel{\rm def}{=}$& $ c(x).\overline{c}(x).I+ d(x).\overline{d}(({\sf fst}(x),ip_A)).I$\\
\end{tabular}
  \caption{Processes for ARAN}\label{aran}
\end{figure}

We assume that each node has received a certification from $T$. The protocol proceeds as follows and we also describe the procedure in GCWN in Fig. \ref{aran}:
\begin{itemize}
  \item The source node $A$ begins the procedure of route to destination $X$ by broadcasting a route require package, $(\{RDP,ip_X,N_A\}_{K_{A-}},[cert_A])$, to its neighbours, where RDP is the package identifier, $ip_X$ is the IP address of the destination $X$, $N_A$ is a nonce, and $cert_A$ is the certification of $A$. See process $A$ in Fig. \ref{aran}.
  \item When a node $B$ receives the message, $msg$, it uses $A$'s public key extracted from $cert_A$ in $msg$ to check the message. If the check fails, the message is dropped; otherwise $B$ sets up a reverse path back to the source by recording the neighbor from which it received the RDP and $B$ signs the received message and appends its certification $cert_B$. We use ${\sf check1}(msg)$ to do this, and use ${\sf getIP}(msg)$ to get the IP address from which the message received. Then $B$ rebroadcasts the message $(\{\{RDP,ip_X,N_A\}_{K_{A-}}\}_{K_{B-}},[cert_A,cert_B])$, built by ${\sf NewMsg1}(msg,ip_B)$. See process $Q$ in Fig. \ref{aran}.
  \item When $B$'s neighbor $C$ receives the message, it checks the message using the certifications of both $A$ and $B$. If the check fails, the message is dropped; otherwise $C$ records $B$ to unicast the reply, removes $B$'s signature and certification, signs the message broadcasted by $A$ and appends its certification. Then $C$ rebroadcasts the message $(\{\{RDP,ip_X,N_A\}_{K_{A-}}\}_{K_{C-}},[cert_A,cert_C])$, constructed by ${\sf NewMsg1}(msg,ip_C)$.
      %$=(\{{\sf getMsg}({\sf fst}(msg), {\sf last}({\sf snd}(msg)))\}_{K_{C-}}, [{\sf head}({\sf snd}(msg)),cert_C])$.
      Each intermediate node along the path repeats the same actions as $C$. See process $Q$ in Fig. \ref{aran}.
  \item When the destination $X$ first receives the RDP, if all the checks are valid then it sends a REP to the source $A$ along the reverse path to the source as a unicast message. $(\{REP,ip_A,N_A\}_{K_{X-}},[cert_X])$ is constructed by ${\sf NewMsg3}(msg,ip_X)$. See process $X$ in Fig. \ref{aran}.
  \item Let $D$ be the first node that receives the REP, $msg$, sent by $X$. After a valid check, node $D$ signs the REP, appends its certification and forwards the message,  $(\{\{REP,ip_A,N_A\}_{K_{X-}}\}_{K_{D-}},[cert_X,cert_D])$ built by ${\sf NewMsg2}(msg,ip_D)$, to the node from which it receives the RDP.
      See process $Q_1$ in Fig. \ref{aran}.
  \item Let $C$ be the next hop of $D$ to $A$. $C$ validates $D$'s signature on the received message, removes $D$'s signature and certification, signs the message and appends its certification, i.e. $(\{\{REP,ip_A,N_A\}_{K_{X-}}\}_{K_{C-}},[cert_X,cert_C])$ constructed by ${\sf NewMsg2}(msg,C)$. Then $C$ unicasts the message to its next hop, i.e. $B$ here.
      Each node along the reverse path repeats the same actions as $C$.
      See process $Q_1$ in Fig. \ref{aran}.
  \item When the source $A$ receives the REP, it validates the destination's signature and the nonce, in a successful state $\overline{\sf s}(0).{\bf 0}$. See process $A_1$ in Fig. \ref{aran}.
\end{itemize}

Here, we implement a unicast using a broadcast, where nodes drop the message if they are not mentioned or addressed.

We have a source $A(ip_A,ip_X)$, a destination $X(ip_X)$, two nodes $B$ and $C$ in the routing path as $Q(ip_B)$ and $Q(ip_C)$, and an intruder $I$ (see Fig. \ref{aran}) which only relays messages. Let the IP address of a node be its location.
Let $N= H\langle \Psi\rangle$, where $H=(\{1,2,3,4\},\{(1,3)\})$, $\Psi(1) = A(1,4)$, $\Psi(2) = Q(2)$, $\Psi(3) = Q(3)$ and $\Psi(4)=X(4)$.
Let $M= G\langle \Phi\rangle$, where $G=(\{1,2,3,4,5\},\{(1,5),(2,5),(4,5),(1,3)\})$, $\Phi(1) = A(1,4)$, $\Phi(2) = Q(2)$, $\Phi(3) = Q(3)$, $\Phi(4)=X(4)$ and $\Phi(5) = I$.
$M$ is an attacked network as a composition of $N$ and $I$.
\begin{figure*}[!t]\small
\setlength{\belowcaptionskip}{-10pt}
  \centering
  \begin{tabular}{|lcll|}
  \hline
  $M$   &$\xrightarrow[]{}$ &$M[1\mapsto A_1(1)][3\mapsto C^{\prime}][5\mapsto\overline{c}(m_1).I]\stackrel{\rm def}{=}M_1$, & by steps in (1) \\
  $M_1$ &$\xrightarrow[]{}\xrightarrow[]{}$  & $M_1[2\mapsto Q_1(2,1)][5\mapsto \overline{c}(m_2).I]\stackrel{\rm def}{=} M_2$, & by steps in (2) \\
  $M_2$ &$\xrightarrow[]{}\xrightarrow[]{}$ &$M_2[5\mapsto \overline{d}({\sf fst}(m_3),1).I][4\mapsto {\bf 0}]\stackrel{\rm def}{=} M_3$, & by steps in (3)\\
  $M_3$ & $\xrightarrow[]{}$ &$M_3[1\mapsto \overline{s}(0).{\bf 0}][5\mapsto I]\stackrel{\rm def}{=} M_4$,& by steps in (4)\\
  \hline
\end{tabular}
  \caption{Reductions in ARAN}\label{transition:aran}
\end{figure*}

Now we show that there is an attack from the intruder $I$ in $M$, by showing that $M$ and $N$ are not weakly barbed bisimilar.
In fact, $M$ can evolve to an incorrect route state through: (1) $A$ broadcasting message $m_1$ to start the routing procedure and $I$ and $C$ receiving the message; (2) $I$ replaying the message to $B$, $B$ rebroadcasting the message $m_2$ signatured by $B$ and only $I$ receiving the message; (3) $I$ rebroadcasting the message $m_2$ to $X$ and $X$ sending a reply $m_3$ to $I$; (4) $I$ sending the replay to $A$ and $A$ reaching an incorrect route state.
See Fig. \ref{transition:aran} for details.
In $M_1$ from Fig. \ref{transition:aran}, $C^{\prime} = \overline{c}({\sf NewMsg1}(m_1,3)).Q_1(3,1)$.
%\begin{center}\tiny
%\begin{tabular}{|lcll|}
%  \hline
%  % after \\: \hline or \cline{col1-col2} \cline{col3-col4} ...
%  $M$   &$\xrightarrow[]{}$ &$M[1\mapsto P_1(ip_A)][3\mapsto C^{\prime}][5\mapsto\overline{c}(m_1).I]\stackrel{\rm def}{=}M_1$, & by steps in (1) \\
%  $M_1$ &$\xrightarrow[]{}\xrightarrow[]{}$  & $M_1[2\mapsto Q_1(ip_B,ip_A)][5\mapsto \overline{c}(m_2).I]\stackrel{\rm def}{=} M_2$, & by steps in (2) \\
%  $M_2$ &$\xrightarrow[]{}\xrightarrow[]{}$ &$M_2[5\mapsto \overline{d}({\sf fst}(m_3),ip_B).I][4\mapsto {\bf 0}]\stackrel{\rm def}{=} M_3$, & by steps in (3)\\
%  $M_3$ & $\xrightarrow[]{}$ &$M_3[1\mapsto \overline{s}(0).{\bf 0}][5\mapsto I]$,& by steps in (4)\\
%  \hline
%\end{tabular}
%\end{center}

In network $N$, node $A$ cannot reach the state $\overline{s}(0).{\bf 0}$. Thus $N$ cannot reach a network $N^{\prime}$ with $N^{\prime}\!\!\downarrow_{\overline{s}}$.
Since $M_4\!\!\downarrow_{\overline{s}}$, $M$ and $N$ are not weakly barbed bisimilar by the definition.
%Similarly, we can also show that $M$ and $N$ are not weakly bisimilar.
\subsection{The Alternating Bit Protocol}
\begin{figure*}[!t]\small
\setlength{\belowcaptionskip}{-10pt}
  \centering
  \begin{tabular}{rcl}
  $P_1(lt_1,b)$ & $\stackrel{\rm def}{=}$ & ${\bf if}~{\sf null}(lt_1)~{\bf then}~\overline{\sf send}((End,b)).{\sf ack}(x).($  \\
   &  &\qquad\qquad\qquad ${\bf if}~ x = (Ack,b)~{\bf then}~{\bf 0}~{\bf else}~P_1(lt_1,b))$ \\
   &  &${\bf else}~\overline{\sf send}(({\sf head}(lt_1),b)).{\sf ack}(x).( $\\
   &  &\qquad\qquad\qquad ${\bf if}~x=(Ack,b)~{\bf then}~P_1({\sf tail}(lt_1),\neg b)~{\bf else}~P_1(lt_1,b))$\\
  $P_2(lt_2,b) $ &$\stackrel{\rm def}{=}$ & ${\sf send}(x).({\bf if}~{\sf snd}(x) = b ~$ \\
   &    & \quad\qquad ${\bf then} ~({\bf if}~{\sf fst}(x) = End~{\bf then}~Succ(lt_2)$\\
   &   & \qquad\qquad\qquad\qquad${\bf else}~\overline{\sf ack}((Ack,b)).P_2({\sf append}(lt_2,{\sf fst}(x)),\neg b)) $ \\
   &  & \quad\qquad ${\bf else}~\overline{\sf ack}(Ack,\neg b).P_2(lt_2,b) )$\\
\end{tabular}
\caption{The Alternating Bit Protocol in GCWN}\label{ABP}
\end{figure*}

The Alternating Bit Protocol (ABP) is a simple data link layer network protocol.
ABP is used when a transmitter $P_1$ wants to send messages to a receiver $P_2$, with the assumptions that the channel may corrupt a message and that $P_1$ and $P_2$ can decide whether they have received a correct message.
%We also assume that the channel from $P_1$ to $P_2$ is initialized. Initially, there are no messages in transit.
Each message from $P_1$ to $P_2$ contains a data part and a one-bit sequence number, i.e. a value that is $0$ or $1$. $P_2$ can send two acknowledge messages, i.e. $(Ack,0)$ and $(Ack,1)$, to $P_1$.

%This ABP example is inspired by \cite{Milner1989,lanese2010operational}.
In \cite{Milner1989}, ABP was formalized in CCS with an interleaving semantics.
In \cite{lanese2010operational}, ABP was investigated in a broadcasting semantics based on transmission radius.
In GCWN, we intend to show that graphs can be used to concisely characterize communicating capacities.

%Alternating Bit Protocol (ABP) is used when a process $P_1$ wants to send a sequence of massages to a target process $P_2$. A message may not be correctly received by $P_2$ in a single transmission. The aim of the protocol is to guarantee that every message is correctly received by $P_2$ in the order of that they are transformed from $P_1$, through the re-transmission of these messages. We consider a simple version of ABP considering erroneous acknowledgements rather than a timeout.

%\begin{center}
%\begin{tabular}{rcl}
%  $P_1(lt_1,b)$ & $\stackrel{\sf def}{=}$ & ${\bf if}~{\sf null}(lt_1)~{\bf then}~\overline{\sf send}((End,b))\cdot({\sf ack}(x)\cdot($  \\
%   &  &\qquad\qquad\qquad ${\bf if}~ x = (Ack,b)~{\bf then}~{\bf 0}~{\bf else}~P_1(lt_1,b)))$ \\
%   &  &${\bf else}~\overline{\sf send}(({\sf head}(lt_1),b))\cdot({\sf ack}(x)\cdot( $\\
%   &  &\qquad\qquad\qquad ${\bf if}~x=(Ack,b)~{\bf then}~P_1({\sf tail}(lt_1),\neg b)~{\bf else}~P_1(lt_1,b)))$\\
%  $P_2(lt_2,b) $ &$\stackrel{\sf def}{=}$ & ${\sf send}(x)\cdot({\bf if}~{\sf snd}(x) = b ~$ \\
%   &    & \quad\qquad ${\bf then} ~({\bf if}~{\sf fst}(x) = End~{\bf then}~Succ(lt_2)$\\
%   &   & \qquad\qquad\qquad\qquad${\bf else}~\overline{\sf ack}((Ack,b))\cdot(P_2({\sf append}(lt_2,{\sf fst}(x)),\neg b))) $ \\
%   &  & \quad\qquad ${\bf else}~\overline{\sf ack}(Ack,\neg b)\cdot(P_2(lt_2,b)) )$\\
%\end{tabular}
%\end{center}
In Fig. \ref{ABP} we provide a specification of ABP in GCWN.
${\sf send}$ and ${\sf ack}$ are channels.
The transmitter $P_1$ has a list $lt_1$ containing the messages to be sent, and the receiver $P_2$ also has a list $lt_2$ containing the received messages. The list is equipped with operations ${\sf head}$ (returning the head of a list), ${\sf tail}$ (returning a list with the first element removed), ${\sf append}$ (inserting an element as the last element of the new list ) and ${\sf null}$ (testing whether a list is empty).
We use ${\sf fst}$ (and ${\sf snd}$) to return the first element (and the second element) of a pair.
$End$ is an entry to indicate that all the messages in $lt_1$ have been transformed. $Succ(lt_2)$ indicates that the receiver has successfully received all the messages. We define a network $M$ consisting of two nodes linked by an edge, and $P_1(lt,b)$ located at one location and $P_2([],b)$ located at the other. Formally,
$M=(\{p_1,p_2\},\{(p_1,p_2)\})\langle\Phi\rangle$, $\Phi(p_1)=P_1(lt,b)$ and $\Phi(p_2)=P_2([],b)$.

\begin{proposition}\label{ABP_prop}
For any network $O$,
$M\oplus O\xrightarrow[]{}^{\ast} M[p_1\mapsto {\bf 0}][p_2\mapsto Succ(lt)]\oplus O^{\prime}$ for some $O^{\prime}$.
\end{proposition}

The proposition says that if the transmitter $P_1$ and the receiver $P_2$ can communicate with each other but they cannot communicate with the nodes in $O$ (i.e. $(M\oplus O)$), then the whole network can reach a state where all the messages in $P_1$ are correctly received by $P_2$ no matter what happens in $O$.
%\begin{IEEEproof}
%Since $O$ does not affect the reductions of $M$, we only consider the reductions of $M$, i.e. interactions between $P_1$ and $P_2$. There are two cases for the reduction
%\begin{itemize}
%  \item either $M\xrightarrow[]{}^{\ast} M[p_1\mapsto P_1(lt_1,b)][p_2\mapsto P_2(lt_2,b)]$ with the concatenation of $lt_1$ and $lt_2$ equals to $lt$, and this is a some stage of the reduction,
%  \item or $M\xrightarrow[]{}^{\ast}M[p_1\mapsto{\bf 0}][p_2\mapsto Succ(lt)]$, and this is the final successful stage of the reduction.
%\end{itemize}
%
%Induction on the length of $lt_1$, we only show some cases and other cases are similar:
%\begin{itemize}
%  \item If ${\sf null}(lt_1)$ is satisfied, then a possible reduction sequence is: sending the $End$ message, receiving acknowledge from $P_2$, passing the conditional evaluation in the sender and in the receiver respectively, then reducing to the final successful stage.
%  \item If ${\sf null}(lt_1)$ is not satisfied, then a possible reduction sequence is: sending the head of $lt_1$, passing the conditional evaluation of the receiver, receiving acknowledge from the receiver, then reducing to the next stage of the reduction.
%\end{itemize}
%\end{IEEEproof}

\section{Related Work}\label{Related_work}
Inspired by \cite{Ehrhard2013ccts,liu2016vccts}, which focus on tree structured concurrent systems with point-to-point communications, we propose a graph-based calculus to study wireless networks with local broadcasts.
Below we only discuss some closely related work on wireless systems.

\noindent{\bf Calculi for Wireless Systems. }
Several process calculi for wireless systems have been proposed.
A brief survey of broadcast calculi can be found in \cite{prasad2014themes}.

%Most related wireless calculi to our work we are aware of are \cite{nanz2006framework,lanese2010operational,ghassemi2008restricted,godskesen2007calculus,singh2010process,merro2009observational,fehnker2012process}.

%In CBS the first broadcast calculus, processes speak one at a time and are heard instantaneously by all others (global broadcast). Only one message can be broadcast at a time. In wireless networks, broadcast have limited and overlapping ranges, these being typically given by a topology separate from the syntax of the processes terms. Wireless networks are not the major area of work, with many different aspects being studied. Wireless calculi are significantly more complex.

CBS$^\sharp$ %(a calculus for mobile wireless networks)
\cite{nanz2006framework}  was probably the first calculus for wireless systems, and it is an extension of CBS \cite{prasad1995calculus}. 
In CBS$^\sharp$, every node is specified by a location, and system transitions are indexed by graphs which represent the connectivity of nodes.
%e.g. $N\xrightarrow[]{(U,m)\sharp}_G N^{\prime}$ meaning that $N$ can evolve to $N^{\prime}$, sending message $U$ by node $m$.
Thus, graphs specify possible behaviors of a system at the semantic level.
Behavioural equivalences are defined to identify processes.
The final goal of CBS$^\sharp$ is to give a framework to specify and analyse communication protocols for wireless networks.
Different from CBS$^\sharp$, in GCWN graphs are introduced at the syntactic level and the weak bisimulation of GCWN takes locations and links into account.

%In CBS$^\sharp$, an extension of CBS, was probably the first calculus for wireless systems. only locations of nodes are specified. so graphs also define the possible behaviours of a system. rely on a graph at semantics level to represent the connectivity of nodes is a separate parameter to the system, and should give correct results for any connectivity. The final goal CBS$^\sharp$ is to given a framework to specify and analyse communication protocols for wireless networks. models a node in a network by an
%$n[p; s]$ notation, specifying process $p$ with store $s$ deployed
%at the logical address of $n$. This is a development of CBS to include varying
%interconnection topologies. Input and output is performed on a universal
%ether and transitions are indexed with topologies which are sets of connectivity graphs; the connectivity graph matters for the input rule (reception
%is possible from any connected location).

CWS (Calculus for Wireless Systems)
\cite{lanese2010operational,mezzetti2006towards} was developed to model protocols at the data-link layer.
In CWS, a node $n[P]_{l,r}^c$ stands for a node named $n$, located at $l$, executing $P$, with channel $c$ and transmission radius $r$.
CWS deals with static topologies, and the topology of a network can be derived by a distance function to compute the nodes in the transmission range of each node. CWS separates the begin and the end of a transmission to handle interferences. The main result of CWS is a correspondence between a reduction semantics and a labelled transition semantics, called harmony theorem. In GCWN, we use graphs to describe connections in networks.
Besides a similar harmony theorem, we also develop behavioural equivalences for GCWN.

%CMN  is also a value-passing calculus.
In CMN (Calculus of Mobile Ad Hoc Networks) \cite{merro2009observational}, the nodes are similar to the ones in CWS.
Both a reduction semantics and a labelled transition semantics are developed, and a harmony theorem is proved for them. The main result is that the labelled bisimilarity coincides with reduction barbed congruence.

CMAN (Calculus for Mobile Ad Hoc Networks)
\cite{godskesen2007calculus} supports local broadcast and dynamic changes of the network topology.
CMAN is equipped with a reduction semantics and
a reduction congruence, and the weak bisimulation coincides with the reduction congruence in CMAN.
CMAN also provides a formalisation of an attack on the cryptographic routing protocol ARAN.
However, the bisimulations in \cite{merro2009observational,godskesen2007calculus} do not take locations and links into account.
And it is unclear how to define a parameterized weak bisimulation in \cite{merro2009observational,godskesen2007calculus}.
%However, the bisimulation does not take locations and links into account either.

Cerone and Hennessy \cite{cerone2014characterising} proposed a calculus for distributed systems,
using directed graphs and equipped with testing preorders in the style of DeNicola and Hennessy \cite{de1984testing}.
Directed graphs are more refined than undirected graphs, but in most situations undirected graphs are enough to specify the connectivity of networks, e.g. \cite{fehnker2012process,godskesen2007calculus,lanese2010operational,merro2009observational,nanz2006framework,singh2010process}.
The testing preorders in \cite{cerone2014characterising} are similar to the barbed congruence in GCWN,
but they only consider actions with the same locations.

\noindent{\bf Calculi for IoT. }
Lanese et al. \cite{lanese2013internet} proposed the first process calculi for Internet of Things (IoT). %with a behavioural equivalence to compare IoT systems. % called IoT-calculus \cite{lanese2013internet}.
%The IoT-calculus captures the partial topology of communications and the interaction between smart devices. A behavioural equivalence is then defined to compare IoT systems.
Recently, a calculus for IoT \cite{lanotte2016semantic} was proposed with a fully abstract semantics in a point-to-point approach.

\noindent{\bf Calculi for CPS. }
Vigo et al. \cite{vigo2013broadcast} proposed a calculus for wireless-based cyber-physical systems (CPSs) to model and reason about cryptographical primitives.
%But it does consider local broadcast and topological structures.
Inspired by \cite{vigo2013broadcast}, Wu and Zhu \cite{wu2015calculus} considered a static network topology and proved a harmony theorem to link reduction semantics and labelled transition semantics.
\section{Conclusions}\label{Conclusions}
We have proposed a graph-based calculus, called GCWN, to model and reason about wireless networks.
In GCWN, we use graphs at syntactical level to specify local broadcast.
The calculus is equipped with a reduction semantics and a labelled transition semantics.
The former has been used to define weak barbed congruence. The latter has been used to define a parameterized weak bisimulation emphasizing
locations and local broadcast.
We have proved that the two semantics model the same behaviours and weak bisimilarity implies weak barbed congruence.
GCWN also has been used to reason about scenarios in ARAN and ABP.

There are some further issues.
Firstly, we are going to extend GCWN to handle dynamic topologies of wireless networks.
%Secondly, we will study the addition of probabilities to our calculus and apply it to wireless systems with noisy channels.
Secondly, we plan to extend GCWN with directed graphs.
Thirdly, we would like to apply our calculus and methods to other wireless-based scenarios, e.g. the ones in IoT and CPSs.
Finally, we plan to develop an implementation of the calculus to assist in reasoning about systems.

\bibliographystyle{IEEEtran}
\bibliography{IEEEabrv,myref}

\begin{thebibliography}{10}
\providecommand{\url}[1]{#1}
\csname url@samestyle\endcsname
\providecommand{\newblock}{\relax}
\providecommand{\bibinfo}[2]{#2}
\providecommand{\BIBentrySTDinterwordspacing}{\spaceskip=0pt\relax}
\providecommand{\BIBentryALTinterwordstretchfactor}{4}
\providecommand{\BIBentryALTinterwordspacing}{\spaceskip=\fontdimen2\font plus
\BIBentryALTinterwordstretchfactor\fontdimen3\font minus
  \fontdimen4\font\relax}
\providecommand{\BIBforeignlanguage}[2]{{%
\expandafter\ifx\csname l@#1\endcsname\relax
\typeout{** WARNING: IEEEtran.bst: No hyphenation pattern has been}%
\typeout{** loaded for the language `#1'. Using the pattern for}%
\typeout{** the default language instead.}%
\else
\language=\csname l@#1\endcsname
\fi
#2}}
\providecommand{\BIBdecl}{\relax}
\BIBdecl

\bibitem{Milner1989}
R.~Milner, \emph{Communication and Concurrency}.\hskip 1em plus 0.5em minus
  0.4em\relax Upper Saddle River, NJ, USA: Prentice-Hall, Inc., 1989.

\bibitem{Hoare1985communicating}
C.~Hoare, \emph{Communicating sequential processes}.\hskip 1em plus 0.5em minus
  0.4em\relax Prentice-hall, 1985.

\bibitem{prasad1995calculus}
K.~V. Prasad, ``A calculus of broadcasting systems,'' \emph{Science of Computer
  Programming}, vol.~25, no.~2, pp. 285--327, 1995.

\bibitem{fehnker2012process}
A.~Fehnker, R.~van Glabbeek, P.~H{\"o}fner, A.~McIver, M.~Portmann, and W.~L.
  Tan, ``A process algebra for wireless mesh networks,'' in \emph{European
  Symposium on Programming}.\hskip 1em plus 0.5em minus 0.4em\relax Springer,
  2012, pp. 295--315.

\bibitem{godskesen2007calculus}
J.~C. Godskesen, ``A calculus for mobile ad hoc networks,'' in
  \emph{International Conference on Coordination Languages and Models}.\hskip
  1em plus 0.5em minus 0.4em\relax Springer, 2007, pp. 132--150.

\bibitem{lanese2010operational}
I.~Lanese and D.~Sangiorgi, ``An operational semantics for a calculus for
  wireless systems,'' \emph{Theoretical Computer Science}, vol. 411, no.~19,
  pp. 1928--1948, 2010.

\bibitem{merro2009observational}
M.~Merro, ``An observational theory for mobile ad hoc networks (full
  version),'' \emph{Information and Computation}, vol. 207, no.~2, pp.
  194--208, 2009.

\bibitem{nanz2006framework}
S.~Nanz and C.~Hankin, ``A framework for security analysis of mobile wireless
  networks,'' \emph{Theoretical Computer Science}, vol. 367, no.~1, pp.
  203--227, 2006.

\bibitem{singh2010process}
A.~Singh, C.~Ramakrishnan, and S.~A. Smolka, ``A process calculus for mobile ad
  hoc networks,'' \emph{Science of Computer Programming}, vol.~75, no.~6, pp.
  440--469, 2010.

\bibitem{Milner1992barbed}
R.~Milner and D.~Sangiorgi, ``Barbed bisimulation,'' in \emph{Automata,
  Languages and Programming}.\hskip 1em plus 0.5em minus 0.4em\relax Springer,
  1992, pp. 685--695.

\bibitem{sanzgiri2005authenticated}
K.~Sanzgiri, D.~LaFlamme, B.~Dahill, B.~N. Levine, C.~Shields, and E.~M.
  Belding-Royer, ``Authenticated routing for ad hoc networks,'' \emph{IEEE
  Journal on selected areas in communications}, vol.~23, no.~3, pp. 598--610,
  2005.

\bibitem{Ehrhard2013ccts}
T.~Ehrhard and Y.~Jiang, ``{CCS} for trees,'' 2013,
  \url{http://arxiv.org/abs/1306.1714}.

\bibitem{liu2016vccts}
S.~Liu and Y.~Jiang, ``Value-passing {CCS} for trees: a theory for concurrent
  systems,'' in \emph{International Symposium on Theoretical Aspects of
  Software Engineering}.\hskip 1em plus 0.5em minus 0.4em\relax {IEEE} Computer
  Society, 2016, pp. 101--108.

\bibitem{boudol1994theory}
G.~Boudol, I.~Castellani, M.~Hennessy, and A.~Kiehn, ``A theory of processes
  with localities,'' \emph{Formal Aspects of Computing}, vol.~6, no.~2, pp.
  165--200, 1994.

\bibitem{Sangiorgi:2011}
D.~Sangiorgi, \emph{Introduction to Bisimulation and Coinduction}.\hskip 1em
  plus 0.5em minus 0.4em\relax New York, NY, USA: Cambridge University Press,
  2011.

\bibitem{prasad2014themes}
K.~Prasad, ``Themes in broadcast calculi,'' in \emph{International Symposium on
  Parallel and Distributed Computing}.\hskip 1em plus 0.5em minus 0.4em\relax
  IEEE, 2014, pp. 16--22.

\bibitem{mezzetti2006towards}
N.~Mezzetti and D.~Sangiorgi, ``Towards a calculus for wireless systems,''
  \emph{Electronic Notes in Theoretical Computer Science}, vol. 158, pp.
  331--353, 2006.

\bibitem{cerone2014characterising}
A.~Cerone and M.~Hennessy, ``Characterising testing preorders for broadcasting
  distributed systems,'' in \emph{International Symposium on Trustworthy Global
  Computing}.\hskip 1em plus 0.5em minus 0.4em\relax Springer, 2014, pp.
  67--81.

\bibitem{de1984testing}
R.~De~Nicola and M.~C. Hennessy, ``Testing equivalences for processes,''
  \emph{Theoretical computer science}, vol.~34, no. 1-2, pp. 83--133, 1984.

\bibitem{lanese2013internet}
I.~Lanese, L.~Bedogni, and M.~Di~Felice, ``Internet of things: a process
  calculus approach,'' in \emph{Proceedings of the 28th Annual ACM Symposium on
  Applied Computing}.\hskip 1em plus 0.5em minus 0.4em\relax ACM, 2013, pp.
  1339--1346.

\bibitem{lanotte2016semantic}
R.~Lanotte and M.~Merro, ``A semantic theory of the internet of things,'' in
  \emph{International Conference on Coordination Languages and Models}.\hskip
  1em plus 0.5em minus 0.4em\relax Springer, 2016, pp. 157--174.

\bibitem{vigo2013broadcast}
R.~Vigo, F.~Nielson, and H.~R. Nielson, ``Broadcast, denial-of-service, and
  secure communication,'' in \emph{International Conference on Integrated
  Formal Methods}.\hskip 1em plus 0.5em minus 0.4em\relax Springer, 2013, pp.
  412--427.

\bibitem{wu2015calculus}
X.~Wu and H.~Zhu, ``A calculus for wireless sensor networks from quality
  perspective,'' in \emph{International Symposium on High Assurance Systems
  Engineering}.\hskip 1em plus 0.5em minus 0.4em\relax IEEE, 2015, pp.
  223--231.

\end{thebibliography}
%\bibliographystyle{IEEEtranS}
%\bibliography{myref}

\newpage
\appendix
\renewcommand{\appendixname}{Appendix~\Alph{section}}

\subsection{Proofs in Section \ref{sec:semantics}}
\begin{lemma}\label{lemma:barb_reductions}
Let $\mathcal{B}$ a weak barbed bisimulation. If $(M,N)\in\mathcal{B}$ and $M\xrightarrow[]{}^{\ast}M^{\prime}$,
then there exists $N^{\prime}$ such that $N\xrightarrow[]{}^{\ast}N^{\prime}$ and $(M^{\prime},N^{\prime})\in \mathcal{B}$.
\end{lemma}
\begin{IEEEproof}
Induction on the length of the derivation $M\xrightarrow[]{}^{\ast}M^{\prime}$.
\end{IEEEproof}

\begin{lemma}\label{lemma:barbedbisimulation}
A symmetric binary relation $\mathcal{B}$ on ${\bf Net}$ is a weak barbed bisimulation if and only if the following conditions hold:
\begin{itemize}
  \item if $(M,N)\in \mathcal{B}$ and $M\xrightarrow[]{}^{\ast}M^{\prime}$, then $N\xrightarrow[]{}^{\ast}N^{\prime}$ and $(M^{\prime},N^{\prime})\in \mathcal{B}$ for some $N^{\prime}$;
  \item if $(M,N)\in \mathcal{B}$ and $M\downarrow_{\overline{c}}$, then $N\xrightarrow[]{}^{\ast}N^{\prime}$ and $N^{\prime}\downarrow_{\overline{c}}$ for some $N^{\prime}$.
\end{itemize}
\end{lemma}
\begin{IEEEproof}
($\Leftarrow$) Because $\xrightarrow[]{}$ is a special case of $\xrightarrow[]{}^{\ast}$, this direction is obvious.

($\Rightarrow$) For the first statement, it is straightforward by Lemma \ref{lemma:barb_reductions}.

For the second statement, it is obvious by the definition of weak barbed bisimulation.
\end{IEEEproof}

Let $\mathcal{B}_1$ and $\mathcal{B}_2$ are binary relations on ${\bf Net}$. Let $\mathcal{B}_2\circ\mathcal{B}_1$ be the composition of $\mathcal{B}_1$ and $\mathcal{B}_2$ such that if $(M,N)\in \mathcal{B}_1$ and $(N,O)\in\mathcal{B}_2$, then $(M,O)\in \mathcal{B}_2\circ\mathcal{B}_1$.

The proof for Lemma \ref{lemma:barbbisim}.
\begin{IEEEproof}
We need to prove that $\stackrel{\bullet}{\approx}$ is reflexive, symmetric and transitive. It is straightforward from the definition and Lemma \ref{lemma:barbedbisimulation}.
\end{IEEEproof}

The proof for Proposition \ref{prop:congruence}.
\begin{IEEEproof}For the first statement,
from the definition of congruence, it is obvious that the identity relation contained in $\mathcal{R}$ is a congruence. And congruences are closed under arbitrary unions and contexts.

For the second statement, let $\mathcal{E}$ be a congruence defined as: $(M,N)\in \mathcal{E}$ if and only if for any context $\mathcal{C}[\cdot]$ one has $(\mathcal{C}[M],\mathcal{C}[N])\in \mathcal{R}$. Therefore, $\mathcal{E}$ is a congruence contained in $\mathcal{R}$ (because we can take $\mathcal{C}[\cdot] = [\cdot]$) and hence $\mathcal{E}\subseteq \overline{\mathcal{R}}$. Conversely, let $(M,N)\in \overline{\mathcal{R}}$ and $\mathcal{C}[\cdot]$ be a context. Because $\overline{\mathcal{R}}$ is a congruence, we have $(\mathcal{C}[M],\mathcal{C}[N])\in \overline{\mathcal{R}}$. We have $(\mathcal{C}[M],\mathcal{C}[N])\in \mathcal{R}$ from $\overline{\mathcal{R}}\subseteq \mathcal{R}$ by definition of $\overline{\mathcal{R}}$ and hence $(M,N)\in \mathcal{E}$.
\end{IEEEproof}

\subsection{Proofs in Section \ref{sec:main_result}}
The proof for Lemma \ref{lamma:equiv}.
\begin{IEEEproof}
We prove it by induction on the depth of the inference of $M\xrightarrow[]{\delta}M^{\prime}$.

First, it is enough to prove the result in the special case that the congruence $M\equiv N$ is due to a single application of a structural congruence rule. And the general case follows just by iterating the special case.

The full proof must treat all possible cases for the final step of the inference of $M\xrightarrow[]{\delta}M^{\prime}$.
Here we only consider one case, and suppose it uses rule (N-ParL), where $M$ is $M_1\oplus_C M_2$, with $M_1\xrightarrow[]{\delta}M_1^{\prime}$ inferred by a shorter inference. Now there are many ways in which $M_1\oplus_C M_2\equiv N$ may be due to a single use of a structural congruence rule, and here we only consider two cases to confirm ourselves.

(1) Suppose that the commutativity rule (i.e. $M\oplus_D N \equiv N\oplus_D M$) is used, so we have $N$ is $M_2\oplus_C M_1$. In this case, we use the rule (N-ParR) to deduce $N\xrightarrow[]{\delta}M_2\oplus_C M_1^{\prime}$. Take $N^{\prime}$ to be $M_2\oplus_C M_1^{\prime}$, and we have $M^{\prime}\equiv N^{\prime}$ as required.

(2) Suppose that a single rule of structural congruence is used in $M_1$, so that $M_1\equiv N_1$ and $N$ is $N_1\oplus_C M_2$. Since $M_1\xrightarrow[]{\delta}M_1^{\prime}$ is inferred by a shorter inference, by applying the induction we have $N_1\xrightarrow[]{\delta}N_1^{\prime}$ and $M_1^{\prime}\equiv N_1^{\prime}$. Take $N^{\prime}$ to be $N_1^{\prime}\oplus_C M_2$, and by using (N-ParL) we can deduce that $N\xrightarrow[]{\delta}N^{\prime}$ and $M^{\prime}\equiv N^{\prime}$ as required.

So the result follows by a fairly lengthy case analysis, both for the structural congruence rule used and for the last step of the transition inference.
\end{IEEEproof}

The proof for Theorem \ref{harmony}.
\begin{IEEEproof}
We first prove that if $M\xrightarrow[]{}M^{\prime}$ then either $M\xrightarrow[]{\tau}M^{\prime\prime}$ and $M^{\prime\prime}\equiv M^{\prime}$ for some $M^{\prime\prime}$, or there is $p:\overline{c}v$ such that $M\xrightarrow[]{p:\overline{c}v}M^{\prime\prime}$ and $M^{\prime\prime}\equiv M^{\prime}$ for some $M^{\prime\prime}$. We prove it by rule induction on the inference of $M\xrightarrow[]{}M^{\prime}$.

Supposed that the transition $M\xrightarrow[]{}M^{\prime}$ is inferred by an application of rule (R-Bcast), that is
$$\tiny \infer[]{M\xrightarrow[]{}M[p\mapsto P][q_i\mapsto Q_i\{v/x\}]_{q_i\in L}}
{M(p)=\overline{c}(e).P + R~~ {\sf eval}(e) = v ~~ L = \{q_i\mid p\frown_M q_i, M(q_i)=c(x_i).Q_{i}+ R_i\}}$$
Then we use the following broadcast transition in Fig. \ref{LTS_network}.
$$\infer[]{M_i\oplus_{D_i} N_i\xrightarrow[]{p:\overline{c}v}M_i^{\prime}\oplus_{D_i}N_i^{\prime}}{M_i\xrightarrow[]{p:\overline{c}v}M_i^{\prime}\quad N_i\xrightarrow[]{q:cv}N_i^{\prime}\quad (p,q)\in D_i}$$
Here, we take $N_i$ as a network that only has one node with location $q_i\in L$, the process on the node is $c(x_i).Q_{i}+ R_i$, and $D_i$ is consistent with the graph of the whole network $M$, i.e. we add $|L|$ networks $N_i$ to the network $M_1$ to obtain the network $M$. And at the beginning, $M_1$ only contains the nodes in $M$ on locations in set $|M|\setminus L$. We can rewrite $M$ as $(\cdots((M_1\oplus_{D_1}N_1)\oplus_{D_2}N_2)\cdots)\oplus_{D_k}N_k$ with $k=|L|$.

And we apply this rule $|L|$ times, such that all the potential receivers receive the broadcasting message.
We get $M\xrightarrow[]{p:\overline{c}v}(\cdots((M_1^{\prime}\oplus_{D_1}N_1^{\prime})\oplus_{D_2}N_2^{\prime})\cdots)\oplus_{D_k}N_k^{\prime}$, and $M^{\prime}$
is $(\cdots((M_1^{\prime}\oplus_{D_1}N_1^{\prime})\oplus_{D_2}N_2^{\prime})\cdots)\oplus_{D_k}N_k^{\prime}$.
Therefore we have $M\xrightarrow[]{p:\overline{c}v}M^{\prime}$ as required.

Suppose that the transition of $M\xrightarrow[]{}M^{\prime}$ is inferred by an application of rule (R-Res), it is similar to the rule (R-Bcast). Moreover, at the last step we use the transition rule (N-Res1), we get $M\xrightarrow[]{\tau}M^{\prime}$ as required.

Suppose that the transition of $M\xrightarrow[]{}M^{\prime}$ is inferred by an application of rule (R-Par), it is similar to the rule (R-Bcast).

Suppose that the transition of $M\xrightarrow[]{}M^{\prime}$ is inferred by an application of rule (N-Struct), that is
$$\infer[]{M \xrightarrow[]{} M^{\prime}}{M\equiv N\quad N\xrightarrow[]{} N^{\prime}\quad N^{\prime}\equiv M^{\prime}}$$
We have that $N\xrightarrow[]{}N^{\prime}$ is a shorter inference. The induction hypothesis tells us that there is $N^{\prime\prime}$ such that $N\xrightarrow[]{\tau}N^{\prime\prime}\equiv N^{\prime}$. Therefore we have $M\xrightarrow[]{\tau}M^{\prime\prime}\equiv M^{\prime}$ by Lemma \ref{lamma:equiv} and the transitivity of $\equiv$. The case for $N\xrightarrow[]{p:\overline{c}v}N^{\prime\prime}\equiv N^{\prime}$ is similar.

Now we prove the converse direction. We consider all the possible cases for the last step of the inference $M\xrightarrow[]{\tau}M^{\prime}$ and $M\xrightarrow[]{p:\overline{c}v}M^{\prime}$.

For $M\xrightarrow[]{p:\overline{c}v}M^{\prime}$, we have $M\xrightarrow[]{p:\overline{c}v}M^{\prime}$ for some $p$, $\overline{c}$ and $v$. By an application of Lemma \ref{equiv_action}, we have
$$M\equiv (M_1\oplus_D M_2)\backslash I$$
and
$$M^{\prime}\equiv (M_1[p\mapsto P][q_i\mapsto Q_i\{v/x_i\}]_{q_i\in L} \oplus_D M_2)\backslash I$$
where ${\sf eval}(e) = v$, $I$ with $c\notin I$, $M(p)=\overline{c}(e).P + R$, $L=\{q_i\mid p\frown_M q_i, M(q_i)= c(x_i).Q_i+R_i\}$, $|M_1|=L\cup\{p\}$, $|M_2|=|M|\setminus |M_1|$ and $D\subseteq |M_1|\times|M_2|$ is consistent with $\frown_M$.

We apply rule (R-Bcast) to $M_1$, then apply rule (R-Par) to $M_1[p\mapsto P][q_i\mapsto Q_i\{v/x\}]_{q_i\in L} \oplus_D M_2$ and apply rule (R-Res) to
$(M_1[p\mapsto P][q_i\mapsto Q_i\{v/x\}]_{q_i\in L} \oplus_D M_2)\backslash I$. Finally, apply rule (R-Struct), we get $M\xrightarrow[]{}M^{\prime}$.

For $M\xrightarrow[]{\tau}M^{\prime}$, suppose the $\tau$-transition is generated by an application of (N-Res1).
So we have $N \xrightarrow[]{p:\overline{c}v}N^{\prime}$ for some $p$, $\overline{c}$, $v$, $M\equiv N\backslash c$
and $M^{\prime}\equiv N^{\prime}\backslash c$.
$N \xrightarrow[]{p:\overline{c}v}N^{\prime}$ can be proved following the previous case.
Then apply rule (R-Struct), we get $M\xrightarrow[]{}M^{\prime}$ as require.

The other cases follow from the congruence rules of the reduction semantics.
\end{IEEEproof}

{\bf Next we prove the Soundness Theorem}

\begin{lemma}\label{lemma:taus}
Let $\mathcal{R}$ be a weak bisimulation. If $(M,E,N)\in \mathcal{R}$ and $M\xrightarrow[]{\tau^{\ast}}M^{\prime}$, then $N\xrightarrow[]{\tau^{\ast}}N^{\prime}$ and $(M^{\prime},E,N^{\prime})\in \mathcal{R}$ for some $N^{\prime}$.
\end{lemma}
\begin{IEEEproof}
Induction on the length of the derivation of $M\xrightarrow[]{\tau^{\ast}}M^{\prime}$.
\end{IEEEproof}
\begin{lemma}\label{lemma:actions}
If $M\xrightarrow[]{\tau^{\ast}}M_1$,
$M_1\xLongrightarrow[]{p:\alpha}M_1^{\prime}$ and $M_1^{\prime}\xrightarrow[]{\tau^{\ast}}M^{\prime}$, then
$M\xLongrightarrow[]{p:\alpha}M^{\prime}$.
\end{lemma}
\begin{IEEEproof}
Straightforward.
\end{IEEEproof}
\begin{lemma}\label{lammebisimulation}%[\bf Localized weak bisimulation]
A symmetric localized relation $\mathcal{R}\subseteq {\bf Net}\times\mathcal{P}({\sf Loc}^2)\times{\bf Net}$ is a weak bisimulation if and only if the following properties hold:
\begin{itemize}
  \item if $(M,E,N)\in \mathcal{R}$ and $M\xLongrightarrow[]{p:\alpha}M^{\prime}$, then there exists $N^{\prime}$ such that  $N\xLongrightarrow[]{q:\alpha}N^{\prime}$, $(p,q)\in E$ and $(M^{\prime},E,N^{\prime})\in \mathcal{R}$;
  \item if $(M,E,N)\in \mathcal{R}$ and $M\xrightarrow[]{\tau^{\ast}}M^{\prime}$, then there exists $N^{\prime}$ such that
      $N\xrightarrow[]{\tau^{\ast}}N^{\prime}$ and $(M^{\prime},E,N^{\prime})\in \mathcal{R}$.
\end{itemize}
\end{lemma}
\begin{IEEEproof}
($\Leftarrow$) Because $\xrightarrow[]{\tau}$ and $\xrightarrow[]{\delta}$ with $\delta\neq \tau$ are special cases of
$\xrightarrow[]{\tau^{\ast}}$ and $\xLongrightarrow[]{\delta}$ respectively, this direction is obvious.

($\Rightarrow$) For the first statement, assume that $(M,E,N)\in \mathcal{R}$ and $M\xLongrightarrow[]{p:\alpha}M^{\prime}$ which is $M\xrightarrow[]{\tau^{\ast}}M_1
\xrightarrow[]{p:\alpha}M_1^{\prime}\xrightarrow[]{\tau^{\ast}}M^{\prime}$, by Lemma \ref{lemma:taus} we can get $N\xrightarrow[]{\tau^{\ast}}N_1$ with $(M_1,E,N_1)\in \mathcal{R}$.

From $M_1\xrightarrow[]{p:\alpha}M_1^{\prime}$ and $(M_1,E,N_1)\in \mathcal{R}$, we can get $N_1\xLongrightarrow[]{q:\alpha}N_1^{\prime}$ with the conditions that $(p,q) \in E$ and $(M_1^{\prime},E,N_1^{\prime})\in \mathcal{R}$.

Since $M_1^{\prime}\xrightarrow[]{\tau^{\ast}}M^{\prime}$ and $(M_1^{\prime}, E,N_1^{\prime})\in \mathcal{R}$,
by Lemma \ref{lemma:taus}, we can have $N_1^{\prime}\xrightarrow[]{\tau^{\ast}}N^{\prime}$ with $(N^{\prime},E,N^{\prime})\in \mathcal{R}$.

With $N\xrightarrow[]{\tau^{\ast}}N_1$, $N_1\xLongrightarrow[]{q:\alpha}N_1^{\prime}$ and $N_1^{\prime}\xrightarrow[]{\tau^{\ast}}N^{\prime}$, by Lemma \ref{lemma:actions} we can get $N\xLongrightarrow[]{q:\alpha}N^{\prime}$.

For the second statement, it is straightforward from Lemma \ref{lemma:taus}.
\end{IEEEproof}
\begin{lemma}[\bf Reflexivity]\label{reflexivity}
Let $\mathcal{I}$ be the localized relation defined by $(M,E,N)\in \mathcal{I}$ if $M=N$ and $E={\rm Id}_{|M|}$. Then $\mathcal{I}$ is a weak bisimulation.
\end{lemma}
\begin{IEEEproof}
Straightforward.
\end{IEEEproof}

Let $\mathcal{R}$ and $\mathcal{S}$ be localized relations. We define a localized relation $\mathcal{S}\circ\mathcal{R}$ for the composition of $\mathcal{R}$ and $\mathcal{S}$. $(M,H,O)\in \mathcal{S}\circ\mathcal{R}$ if $H\subseteq |M|\times|O|$ and there exist $N$, $E$ and $F$ such that $(M,E,N)\in \mathcal{R}$, $(N,F,O)\in \mathcal{S}$ and $F\circ E\subseteq H$. Let $F\circ E =\{(p,r)\mid (p,q)\in E, (q,r)\in F\}$.
\begin{lemma}[\bf Transitivity]\label{transitivity}
If $\mathcal{R}$ and $\mathcal{S}$ are weak bisimulations, then $\mathcal{S}\circ\mathcal{R}$ is a weak bisimulation.
\end{lemma}
\begin{IEEEproof}
Obviously, $\mathcal{S}\circ\mathcal{R}$ is symmetric.
Then the proof just follows the definition of the weak bisimulation using the Lemma \ref{lammebisimulation}.

From the hypothesis, let $(M,E,N)\in \mathcal{R}$, $(N,F,O)\in \mathcal{S}$ and $(M,H,O)\in\mathcal{S}\circ\mathcal{R}$ with $F\circ E \subseteq H$.

(1) If $M\xLongrightarrow[]{p:\alpha}M^{\prime}$, then $N\xLongrightarrow[]{q:\alpha}N^{\prime}$, $(p,q)\in E$
 and $(M^{\prime},E,N^{\prime})$ $\in \mathcal{R}$.
From $(N,F,O)\in \mathcal{S}$ and $N\xLongrightarrow[]{q:\alpha}N^{\prime}$, we have $O\xLongrightarrow[]{r:\alpha}O^{\prime}$, $(q,r)\in F$ and $(N^{\prime},F,O^{\prime})\in \mathcal{S}$.
Therefore, for any pair of labels $p:\alpha$ and $q:\alpha$ and pair of labels $q:\alpha$ and $r:\alpha$, we have $(p,r)$ $\in F\circ E\subseteq H$.
So we have $(M^{\prime}, F\circ E, O^{\prime})\in \mathcal{S}\circ \mathcal{R}$.

(2) From $(M,E,N)\in \mathcal{R}$, if $M\xrightarrow[]{\tau^{\ast}}M^{\prime}$, then we have $N\xrightarrow[]{\tau^{\ast}}N^{\prime}$ and $(M^{\prime},E,N^{\prime})\in \mathcal{R}$.
Since $(N,F,O)\in \mathcal{S}$ and $N\xrightarrow[]{\tau^{\ast}}N^{\prime}$, we have
$O\xrightarrow[]{\tau^{\ast}}O^{\prime}$ and $(N^{\prime},F,O^{\prime})\in \mathcal{S}$.
We just get $(M^{\prime},F\circ E, O^{\prime})\in \mathcal{S}\circ \mathcal{R}$ as required.
\end{IEEEproof}

Proof for Lemma \ref{lemma:bisim_equiv}.
\begin{IEEEproof}
%$\approx$ is symmetric by definition. We need to prove $\approx$ is reflexive and transitive, see Appendix for details.
Because $\approx$ is reflexive by Lemma \ref{reflexivity} , symmetric from the definition and transitive by Lemma \ref{transitivity}.
\end{IEEEproof}

The proof for Proposition \ref{propositionparallel}.
\begin{IEEEproof}
From the definitions of adapted relations and parallel extension, it is straightforward to show that $\mathcal{R}^{\prime}$ is symmetric.

Let $(U,F,V)\in \mathcal{R}^{\prime}$ with $(M,E,N)\in \mathcal{R}$, $U= O\oplus_C M$, $V= O\oplus_D N$, $(C,D,E)$ is adapted and $F=\mbox{Id}_{|O|}\cup E$.

\textit{\textbf{Case of a sending transition.}} Given $U\xrightarrow[]{p:\overline{c}v}U^{\prime}$ with $p\in |U|$, we have to show
$V\xLongrightarrow[]{q:{\overline c}v}V^{\prime}$ with $q\in |V|$, $(p,q)\in F$ and $(U^{\prime},F,V^{\prime})\in \mathcal{R}^{\prime}$.
There are two cases for the sending transition from $U=O\oplus_C M$.

(1) The sending transition occurs in $O$, and $O\oplus_C M\xrightarrow[]{p:{\overline c}v}O^{\prime}\oplus_{C}M$.
So, we have $p\in |O|$ and $O\xrightarrow[]{p:{\overline c}v}O^{\prime}$.
Then we have $U^{\prime}=O^{\prime}\oplus_{C} M$.

Similarly, for $V=O\oplus_D N$ we have $V\xrightarrow[]{p:{\overline c}v}V^{\prime}=O^{\prime}\oplus_{D}N$. By the definition of $F$, we have $(p,p)\in {\rm Id}_{|O|}$, i.e. $(p,p)\in F$.
Then it is obvious that the triple $(C,D,E)$ is adapted.
Therefore, we have $(U^{\prime},F,V^{\prime})\in \mathcal{R}^{\prime}$ and $F = \mbox{Id}_{|O|}\cup E$, as required.

(2) The sending transition occurs in $M$ and $O\oplus_C M\xrightarrow[]{p:{\overline c}v} O\oplus_{C}M^{\prime}$.
So we have $p\in |M|$ and $M\xrightarrow[]{p:{\overline c}v}M^{\prime}$, i.e. $U^{\prime} = O\oplus_{C} M^{\prime}$.

Since $(M,E,N)\in \mathcal{R}$, from $M\xrightarrow[]{p:{\overline c}v}M^{\prime}$, we have $N\xLongrightarrow[]{q:{\overline c}v}N^{\prime}$, $(p,q)\in E$
and $(M^{\prime},E,N^{\prime})\in \mathcal{R}$.
We can decompose $N\xLongrightarrow[]{q:{\overline c}v}N^{\prime}$ as
$$ N\xrightarrow[]{\tau^{\ast}}N_1\xrightarrow[]{q:{\overline c}v}N_1^{\prime}\xrightarrow[]{\tau^{\ast}}N^{\prime}$$
We have $V\xrightarrow[]{\tau^{\ast}}V_1$ with $V_1=O\oplus_{D}N_1$.

Similarly, $V\xLongrightarrow[]{q:{\overline c}v}V^{\prime}$
can be decomposed as
$$O\oplus_D N \xrightarrow[]{\tau^{\ast}}O\oplus_{D}N_1 \xrightarrow[]{q:{\overline c}v}
O\oplus_{D}N_1^{\prime}\xrightarrow[]{\tau^{\ast}}O\oplus_{D}N^{\prime}$$

We have $V^{\prime}= O\oplus_{D}N^{\prime}$.
Since $F\subseteq |U|\times |V|$ and $F= \mbox{Id}_{|O|}\cup E$, we have $(p,q)\in F$.
Moreover, the triple $(C,D,E)$ is adapted.
Therefore, we have $(U^{\prime},F,V^{\prime})\in \mathcal{R}^{\prime}$ and $F = \mbox{Id}_{|O|}\cup E$, as required.

\textit{\textbf{Case of a receiving transition.}} Given $U\xrightarrow[]{p:cv}U^{\prime}$ with $p\in |U|$, we have to show
$V\xLongrightarrow[]{q:cv}V^{\prime}$ with $q\in |V|$, $(p,q)\in F$ and $(U^{\prime},F,V^{\prime})\in \mathcal{R}^{\prime}$.
There are two cases for the receiving transition from $U=O\oplus_C M$. The analysis is similar to the case of a sending transition.

\textit{\textbf{Case of a $\tau$-transition.}} Given $U\xrightarrow[]{\tau}U^{\prime}$, we have to show $V\xrightarrow[]{\tau^{\ast}}V^{\prime}$ and $(U^{\prime},F,V^{\prime})\in \mathcal{R}^{\prime}$.
There are two cases for a $\tau${\it -transition} from $U=O\oplus_C M$.
Because $\tau$-transitions are obtained by restriction rules from outputs, i.e. there exist $U\equiv U_1\backslash I$ and
$ U_1\xrightarrow[]{p:{\overline c}v}U_1^{\prime}$ with $p\in |U|=|U_1|$. Then with an application of rule (N-Res1), we get $ U\xrightarrow[]{\tau}U^{\prime}$ and $U^{\prime}\equiv U_1^{\prime}\backslash I$.
Then we have to analysis the cases for
$ U_1\xrightarrow[]{p:{\overline c}v}U_1^{\prime}$ with $p\in |U|=|U_1|$, and this is similar to the case of a sending transition.
\end{IEEEproof}

The proof for Theorem \ref{bisimilationCongruence}.
\begin{IEEEproof}
$\approx$ is an equivalence by Lemma \ref{lemma:bisim_equiv}.
Here, we just need to prove that if $M$ and $N$ are two networks, and $M\approx N$, (i.e. $(M,E,N)\in \mathcal{R}$ for some weak bisimulation $\mathcal{R}$), then
\begin{itemize}
  \item[(1)] $\mathcal{R}$'s parallel extension is a weak bisimulation,
  \item[(2)] $(M\backslash c, E, N\backslash c)$ is contained in some weak bisimulation for any channel $c$.
\end{itemize}

For the proof of (1), we directly apply the Proposition \ref{propositionparallel}.

For the proof of (2), it is sufficient to show that the localized relation
$$\mathcal{S}\stackrel{\rm def}{=}\{((M\backslash c, E, N\backslash c))\mid (M,E,N)\in \mathcal{R} \mbox{ for any channel } c\}$$
is a weak bisimulation.
It is obvious that $\mathcal{S}$ is symmetric.
Then we do a case analysis on the possible transition from $M\backslash c$. The proof is straightforward.
\end{IEEEproof}

Proof for Proposition \ref{propsitionbisimulationbarb}.
\begin{IEEEproof}
Let $\mathcal{B}$ be a binary relation on networks defined by: $(M,N)\in \mathcal{B}$ if $M\approx N$. Then we have to prove that $\mathcal{B}$ is a weak bared bisimulation. First, we know that $\mathcal{B}$ is symmetric, because $\approx$ is symmetric. Then we need to prove $\mathcal{B}$ is reduction closed and barb preserving.

(1) Let $(M,N)\in \mathcal{B}$. If $M\xrightarrow[]{}M^{\prime}$ which is $M\xrightarrow[]{p:\overline{c}v}M^{\prime\prime}\equiv M^{\prime}$ by Theorem \ref{harmony}.
Because $M\approx N$,
we have $N\xLongrightarrow[]{q:\overline{c}v}N^{\prime}$ (i.e. $N\xrightarrow[]{}^{\ast}N^{\prime}$ by by Theorem \ref{harmony}) and $M^{\prime\prime}\approx N^{\prime}$.
We also have $M^{\prime\prime}\approx M^{\prime}$ by Proposition \ref{equiv_bisim}. Since $\approx$ is an equivalence, we have $M^{\prime}\approx N^{\prime}$.
Thus, we have $(M^{\prime},N^{\prime})\in \mathcal{B}$.

(2) Let $(M,N)\in \mathcal{B}$. If $M\downarrow_{\overline c}$,
then there exists a transition $M\xrightarrow[]{p:\overline{c}v}M^{\prime}$. Since $M\approx N$,
we have $N\xLongrightarrow[]{q:\overline{c}v}N^{\prime}$ and $M^{\prime}\approx N^{\prime}$.
$N\xLongrightarrow[]{q:\overline{c}v}N^{\prime}$ means $N\xrightarrow[]{\tau^{\ast}}N_1$ (i.e. $N\xrightarrow[]{}^{\ast}N_1$ by Theorem \ref{harmony})
with $N_1\downarrow_{\overline c}$ for some $N_1$. From $M\downarrow_{\overline c}$, we get that
$N\rightarrow^{\ast}N_1$ with $N_1\downarrow_{\overline c}$ as required.
\end{IEEEproof}

\subsection{Proofs in Section \ref{sec:case_study}}
The proof for Proposition \ref{ABP_prop}.
\begin{IEEEproof}
Since $O$ does not affect the reductions of $M$, we only consider the reductions of $M$, i.e. interactions between $P_1$ and $P_2$. There are two cases for the reduction
\begin{itemize}
  \item either $M\xrightarrow[]{}^{\ast} M[p_1\mapsto P_1(lt_1,b)][p_2\mapsto P_2(lt_2,b)]$ with the concatenation of $lt_1$ and $lt_2$ equals to $lt$, and this is a some stage of the reduction,
  \item or $M\xrightarrow[]{}^{\ast}M[p_1\mapsto{\bf 0}][p_2\mapsto Succ(lt)]$, and this is the final successful stage of the reduction.
\end{itemize}

Induction on the length of $lt_1$, we only show some cases and other cases are similar:
\begin{itemize}
  \item If ${\sf null}(lt_1)$ is satisfied, then a possible reduction sequence is: sending the $End$ message, receiving acknowledge from $P_2$, passing the conditional evaluation in the sender and in the receiver respectively, then reducing to the final successful stage.
  \item If ${\sf null}(lt_1)$ is not satisfied, then a possible reduction sequence is: sending the head of $lt_1$, passing the conditional evaluation of the receiver, receiving acknowledge from the receiver, then reducing to the next stage of the reduction.
\end{itemize}
\end{IEEEproof}

\end{document}